
\hfuzz=5pt
\def\underarrow#1{\mathrel{\mathop{\longrightarrow}\limits_{#1}}}
\def\Tr{\,{\rm Tr}\,}
\def\tr{\,{\rm tr}\,}
\def\upleftrightarrow#1{\raise1.5ex\hbox{$\leftrightarrow$}\mkern-16.5mu #1}
\def\ll{\left\langle}
\def\rr{\right\rangle}
\def\thru#1{\mathrel{\mathop{#1\!\!\!/}}}
\def\square{\kern1pt\vbox{\hrule height 1.2pt\hbox{\vrule width 1.2pt\hskip 3pt
   \vbox{\vskip 6pt}\hskip 3pt\vrule width 0.6pt}\hrule height 0.6pt}\kern1pt}
\magnification=1200
\hoffset=-.1in
\voffset=-.2in
\vsize=7.5in
\hsize=5.6in
\tolerance 10000

\baselineskip 12pt plus 1pt minus 1pt
\pageno=0
\centerline{\bf CONFORMAL SYMMETRY AND DIFFERENTIAL}
\medskip
\centerline{{\bf REGULARIZATION OF THE THREE-GLUON VERTEX}\footnote{*}{This
work is supported in part by funds provided by the U.S. Department of Energy
(D.O.E.) under contract \#DE-AC02-76ER03069, the National Science Foundation
under grant \#87-08447 and CICYT (Spain) under grant \#AEN90-0040.}}
\bigskip
\centerline{Daniel Z. Freedman\footnote{$^1$}{e-mail: dzf@math.mit.edu}}
\vskip 12pt
\centerline{\it Department of Mathematics}
\centerline{and}
\centerline{\it Center for Theoretical Physics}
\centerline{\it Massachusetts Institute of Technology}
\centerline{\it Cambridge, Massachusetts\ \ 02139\ \ \ U.S.A.}
\vskip 12pt
\centerline{Gianluca Grignani,\footnote{$^2$}{On leave from
Dipartimento di Fisica, Universit\'a di Perugia and INFN, Sezione di Perugia,
06100 Perugia, ITALY. e-mail: grignani@ipginfn}
\ Kenneth Johnson\footnote{$^3$}{e-mail:
knjhnsn@mitlns.bitnet}\ and\ Nuria
Rius\footnote{$^4$}{e-mail: rius@mitlns.bitnet}}
\vskip 12pt
\centerline{\it Center for Theoretical Physics}
\centerline{\it Laboratory for Nuclear Science}
\centerline{\it and Department of Physics}
\centerline{\it Massachusetts Institute of Technology}
\centerline{\it Cambridge, Massachusetts\ \ 02139\ \ \ U.S.A.}
\vskip 1.5in
\centerline{Submitted to: {\it Annals of Physics\/}}
\vfill
\centerline{ Typeset in $\TeX$ by Roger L. Gilson}
\vskip -12pt
\noindent CTP\#1991\hfill March 1992
\eject
\baselineskip 24pt plus 1pt minus 1pt
\centerline{\bf ABSTRACT}
\medskip
The conformal symmetry of the QCD Lagrangian for massless quarks is broken
both by renormalization effects and the gauge fixing procedure.  Renormalized
primitive divergent amplitudes have the property that their form away from
the overall coincident point singularity is fully determined by the bare
Lagrangian, and scale dependence is restricted to $\delta$-functions at the
singularity.  If gauge fixing could be ignored, one would expect these
amplitudes to be conformal invariant for non-coincident points.
We find that the one-loop three-gluon vertex function
$\Gamma_{\mu\nu\rho}(x,y,z)$ is conformal invariant in this sense, if
calculated in the background field formalism using the Feynman gauge for
internal gluons.  It is not yet clear why the expected breaking due to gauge
fixing is absent.  The conformal property implies that the gluon, ghost
and quark loop contributions to $\Gamma_{\mu\nu\rho}$ are each purely
numerical combinations of two universal conformal tensors
$D_{\mu\nu\rho}(x,y,z)$ and $C_{\mu\nu\rho}(x,y,z)$ whose explicit form is
given in the text.  Only $D_{\mu\nu\rho}$ has an ultraviolet divergence,
although $C_{\mu\nu\rho}$ requires a careful definition to resolve the
expected ambiguity of a formally linearly divergent quantity.  Regularization
is straightforward and leads to a renormalized vertex function which satisfies
the required Ward identity, and from which the beta-function is easily
obtained.  Exact conformal invariance is broken in higher-loop orders, but we
outline a speculative scenario in which the perturbative structure of the
vertex function is determined from a conformal invariant primitive core by
interplay of the renormalization group equation and Ward identities.

Other results which are relevant to the conformal property include the
following:
\medskip
\item{1)}An analytic calculation shows that the linear deviation from the
Feynman gauge is not conformal invariant, and a separate computation using
symbolic manipulation confirms that among $D_\mu b_\mu$ background gauges,
only the Feynman gauge is conformal invariant.
\medskip
\item{2)}The conventional ({\it i.e.\/} non-background) gluon vertex function
is not conformal invariant because the Slavnov--Taylor identity it satisfies
is more complicated than the simple Ward identity for the background vertex,
and a superposition of $D_{\mu\nu\rho}$ and $C_{\mu\nu\rho}$ necessarily
satisfies a simple Ward identity.  However, the regulated conventional vertex
can be expressed as a multiple of the tensor $D_{\mu\nu\rho}$ plus an
ultraviolet finite non-conformal remainder.  Mixed vertices with both external
background and quantum gluons have similar properties.
\vfill
\eject
\noindent{\bf I.\quad INTRODUCTION}
\medskip
\nobreak
The differential regularization procedure$^1$ gives
a simple, practical method for calculation of the renormalization
group functions and the explicit forms of correlation functions in massless
$\varphi^4$ theory.  The supersymmetric Wess--Zumino model appears equally
simple to treat by this method,$^2$ and complete calculations have been done
in these theories through three-loop order.

A problematic feature of the procedure emerged in the various one-loop
calculations for gauge theories presented in Ref.~[1].  Ward identities must
be studied explicitly to fix the various mass scales which are the parameters
of the regularization scheme.  For example in massless quantum electrodynamics
the renormalized electron vertex function and self-energy are (in the notation
of Sections~II.C and II.B of Ref.~[1])
$$\eqalign{
V_\lambda( x,y,z) &= - 2 \gamma_b \gamma_\lambda \gamma_a V_{ab} (x-z,y-z) \cr
V_{ab} (x,y) &= \left( {\partial\over\partial x_a} + {\partial\over \partial
y_a}\right) \left[ {1\over x^2} \  {\partial\over\partial y_b}\left( {1\over
y^2}\right) {1\over (x-y)^2}\right] \cr
&\qquad - {1\over x^2}\ {1\over (x-y)^2} \left( {\partial\over\partial y_a}\
{\partial\over\partial y_b} - {1\over 4} \delta_{ab} \square_y\right) {1\over
y^2} - {\pi^2\over 4}\delta_{ab} \delta(y) \square {\ln M^2_V x^2\over x^2}
\cr
\Sigma(x) &= {1\over 4} \thru\partial \square {\ln M^2_\Sigma x^2\over x^2}\ \
\ \ ,\cr}\eqno(1.1)$$
The Ward identity
$${\partial\over\partial z_\lambda} V_\lambda (x,y,z) =
\left[\delta(z-x) - \delta(z-y)\right]\Sigma (x-y) \eqno(1.2)$$
is violated (by terms proportional to $\left[ \delta(z-x) - \delta (z-y)\right]
\thru\partial \delta(x-y)$) unless the mass scales are chosen to satisfy $\ln
\left( M_\Sigma/M_V\right) = 1/4$, and it is not difficult to demonstrate
this.

The three-gluon vertex is a fundamental correlation function of non-Abelian
gauge theories.  It is linearly divergent by power counting and provides a
test of the compatibility of differential regularization with Ward identities
in a more singular situation than previously explored.  In this note we report
one-loop results for the three-gluon vertex.  These results are quite simple,
because of the somewhat surprising property that the bare amplitude in real
space is conformal invariant (if calculated in a special gauge).  It is not
clear whether the conformal property is relevant beyond one-loop and whether
simple calculations of higher loop contributions are possible.  However,
one can outline a certain conformal scenario based on the combined constraints
of renormalization group equations and Ward identities on these questions.

The first aspect of the three-gluon vertex we considered were the Ward
identities it must satisfy.  If the vertex is calculated by standard methods
in a covariant gauge, then the analogue of (1.2) is a complicated
Slavnov--Taylor identity involving not only the divergence of the vertex  and
the gluon self-energy, but also the vertex with external ghost lines and ghost
self-energy which enter non-linearly (see Section~2.5 of Ref.~[3] for the
explicit form of this Slavnov--Taylor identity).  Fortunately, the structure
of the Ward identities is far simpler if one-particle irreducible amplitudes
are calculated in the background field formalism developed for gauge theories
by DeWitt,$^4$ 't~Hooft,$^5$ and Abbott,$^6$ because the generating functional
is invariant under gauge transformations of the background.  We therefore use
the background field method as the basis for our work.  Although we work only
to one-loop order, the method is quite general, and it is known$^{7,\,8}$ that
the correct $S$-matrix is obtained when 1PI amplitudes are assembled in tree
structures.  The background field method is equivalent to ordinary field
theory in a special gauge.$^6$

Let us now discuss conformal invariance and its effect in our work.  Conformal
transformations in $d$-dimensional Euclidean space may be defined as the
transformation of points given by
$$x_\mu \to x'_\mu = {x_\mu + c_\mu x^2\over
1 + 2c\cdot x+c^2x^2}\ \ ,\eqno(1.3)$$
where $c_\mu$ is a constant vector.  One can show$^9$ that conformal and
Lorentz invariance imply invariance under the discrete inversion $x_\mu\to
x'_\mu = x_\mu/x^2$, and that the transformation (1.3) can be described as
inversion followed by translation by $c_\mu$ followed by a second inversion.
The full group containing conformal, scale, and Lorentz transformations plus
translations is ${\cal O}(d+1,1)$.

If the correlation functions of a quantum field theory were conformal
invariant, then the spatial dependence of two- and three-point functions would
be almost completely fixed.  Consider, for example, conserved currents
$J^a_\mu(x)$ of scale dimension three in $d=4$ dimensions which obey the
current commutation relations of a Lie algebra with structure constants
$f^{abc}$.  Then covariant two- and three-point functions obey the Ward
identities
$${\partial\over\partial z_\rho} \ll J^a_\mu(x) J^b_\mu(y) J^c_\rho(z)\rr = -
f^{cad} \delta(z-x) \ll J^d_\mu (x) J^b_\nu (y)\rr - f^{cbd} \delta(z-y) \ll
J^a_\mu(x) J^d_\nu(y)\rr\ \ .\eqno(1.4)$$
With these assumptions, one can follow Schreier$^9$ who uses the inversion
property
$$J^{'a}_\mu\left({x_\sigma\over x^2}\right)
= x^6\left( \delta_{\mu\nu} - {2x_\mu
x_\nu\over x^2}\right) J^a_\nu \left(x_\sigma\right) \eqno(1.5)$$
to show that the two-point function, if conformal invariant, must take the
gauge invariant form
$$\eqalign{
\ll J^a_\mu(x) J^b_\nu(y)\rr &= - {1\over 2} k \,\delta^{ab} {1\over (x-y)^4}
{\partial\over\partial x_\mu}\ {\partial\over\partial y_\nu} \ln (x-y)^2 \cr
&= k\,\delta^{ab} \left[ {\delta_{\mu\nu}\over (x-y)^6} -
{2(x-y)_\mu(x-y)_\nu\over(x-y)^8}\right]\cr
&= - {1\over 12} k \delta^{ab} \left( {\partial\over\partial x_\mu}\
{\partial\over\partial x_\nu} - \delta_{\mu\nu}\square \right) {1\over
(x-y)^4} \cr}\eqno(1.6)$$
where $k$ is a constant, while conformal invariant three-point functions
must be linear combinations of two possible conformal tensors
$$\ll J^a_\mu (x) J^b_\nu(y) J^c_\rho(z)\rr = f^{abc} \left( k_1 D^{\rm
sym}_{\mu\nu\rho} (x,y,z) + k_2 C^{\rm sym}_{\mu\nu\rho} (x,y,z)\right)
\eqno(1.7)$$
where $D^{\rm sym}_{\mu\nu\rho}(x,y,z)$ and $C^{\rm sym}_{\mu\nu\rho} (x,y,z)$
are permutation odd tensor functions, obtained from the specific tensors
$$\eqalignno{D_{\mu\nu\rho}(x,y,z) &= {1\over (x-y)^2(z-y)^2(x-z)^2} \
{\partial\over \partial x_\mu}\  {\partial\over\partial y_\nu} \ln (x-y)^2
{\partial\over\partial z_\rho} \ln \left( {(x-z)^2\over (y-z)^2} \right) &(1.8)
\cr
C_{\mu\nu\rho}(x,y,z) &= {1\over (x-y)^4} \ {\partial\over \partial x_\mu}\
{\partial\over\partial z_\alpha} \ln (x-z)^2 {\partial\over\partial y_\nu} \
{\partial\over\partial z_\alpha} \ln (y-z)^2 {\partial\over\partial z_\rho}
\ln \left( {(x-z)^2\over (y-z)^2}\right) &(1.9)\cr}$$
by adding cyclic permutations
$$\eqalign{D^{\rm sym}_{\mu\nu\rho} (x,y,z) &= D_{\mu\nu\rho}(x,y,z) +
D_{\nu\rho\mu} (y,z,x) + D_{\rho\mu\nu} (z,x,y) \cr
C^{\rm sym}_{\mu\nu\rho} (x,y,z) &= C_{\mu\nu\rho} (x,y,z) + C_{\nu\rho\mu}
(y,z,x) + C_{\rho\mu\nu} (z,x,y) \ \ .\cr}\eqno(1.10)$$
(Although not required for a first reading of this paper, we note that only
four of the six tensors which appear in (1.10) are linearly independent since
the combination $D_{\mu\nu\rho} (x,y,z)+{1\over 2} C_{\mu\nu\rho}(x,y,z)$ is
cyclically symmetric.  A convenient basis, equivalent to that of Schreier, is
given by $C^{\rm sym}_{\mu\nu\rho}(x,y,z)$ and $D_{\mu\nu\rho}(x,y,z)$,
$D_{\nu\rho\mu}(y,z,x)$, $D_{\rho\mu\nu}(z,x,y)$.  There are no permutation
even combinations of this basis, so that the $d$-symbol $d^{abc}$ cannot
appear in (1.7).
It will be seen later that $k_1$ and $k$ are related by the Ward identity
(1.4),
while $k_2$ is an independent constant.)

Many readers may now think that these considerations are irrelevant to the
three-gluon vertex and even suspicious, because it is well-known that
correlation functions in massless four-dimensional renormalizable field
theories are not conformal invariant.  Invariance fails because a scale is
introduced in the renormalization procedure in (virtually) all such theories,
while in gauge theories conformal invariance is also broken by the gauge
fixing procedure.  It turns out that the first difficulty is easier to
explain than the second, at least in one-loop order.  It is true that
conformal invariance fails because of renormalization, but in real space at
one-loop order renormalization affects only singular $\delta$-functions
$\delta(x-y)\delta(y-z)$, while for non-coincident points the renormalized
and bare amplitudes coincide.  Thus real space one-loop amplitudes can well be
conformal invariant away from short distance singularities.

Indeed Baker and Johnson$^{10}$ considered the three-point current correlation
function in a theory containing spinor doublet $\psi_i$ with Abelian gauge
interactions.  The ``triangle function'' of the $SU(2)$ current operator
$J^a_\mu = {1\over 2} \bar{\psi}_i \tau^a_{ij} \gamma_\mu \psi_j$ was
shown to have the conformal structure (1.7 -- 1.10)
not only in one-loop, where it
is fairly trivial, but also in two-loop order where it vastly simplified the
calculation.  Conformal invariance held at the two-loop level, because
subdivergences cancel due to the Abelian Ward identity $Z_1=Z_2$, so
renormalization again has no effect for non-coincident points.

The idea that amplitudes away from singularities have the conformal symmetry
of the bare Lagrangian is not sufficient to explain a conformal structure for
the three-gluon vertex function, because the gauge fixing terms in the
Lagrangian break conformal invariance.  Indeed, the gluon propagator does not
transform$^{10}$ as expected from the formal inversion property
$$A'_\mu\left({x_\sigma\over x^2}\right)
=  x^2 \left( \delta_{\mu\nu} - {2x_\mu x_\nu\over
x^2} \right) A_\mu \left( x_\sigma\right)\ \ .\eqno(1.11)$$
However, one can show that the one-loop gluon vertex and self-energy in the
background field formalism satisfy Ward identities of the same simple form as
(1.4), and further that the quark, Faddeev--Popov ghost, and gluon
contributions
satisfy the Ward identities separately.  The quark triangle function is
independent of the background method and the same in one-loop order as in the
Abelian theory above and thus conformal invariant.  Our calculations then
showed that the ghost triangle function was conformal invariant, while the
gluon triangle and seagull graphs combine into an amplitude with the conformal
structure (1.7 -- 1.10)
if the Feynman gauge is used for internal gluons.  Separate
computations show that the background field vertex is
not conformal invariant away from the Feynman gauge, and it is easy to see
that the three-gluon
vertex in ordinary field theory in the Feynman gauge is not conformal
invariant.  Thus the conformal property appears to be very specialized, and,
apart from the calculations themselves, we do not have any qualitative
explanation for it.  For example, a functional identity which expresses the
conformal variation of the generating functional does not suggest that it
vanishes without detailed calculation.

Whether by accident or part of Nature's design, the fact that quark, ghost,
and virtual gluon contributions to the vertex are all numerical combinations
of the invariant tensors $D^{\rm sym}_{\mu\nu\rho}$ and $C^{\rm
sym}_{\mu\nu\rho}$ vastly simplifies the task of regularization.
The bare amplitude $D^{\rm sym}_{\mu\nu\rho}$ has an ultraviolet divergent
Fourier transform, but it is easily regulated using the ideas of differential
regularization.  The story of the tensor $C^{\rm sym}_{\mu\nu\rho}$ is
slightly more involved.  Although formally linearly divergent, its Fourier
transform is ultraviolet finite but subject to shift ambiguities similar
to those of the fermion triangle anomaly.  A shift term changes $C^{\rm
sym}_{\mu\nu\rho}$ by a linear polynomial in momenta which is proportional to
the bare Yang--Mills vertex (see (1.12)).  Regularization is required to
specify the ambiguity in $C^{\rm sym}_{\mu\nu\rho}$, and consideration of
conformal invariance and differential regularization lead to a simple
regulated form which contributes trivially to the Ward identity (1.4).  The
result of these considerations is that the regulated form of $D^{\rm
sym}_{\mu\nu\rho}$ alone determines both the Ward identity and the
$\beta$-function.
The quark, ghost, and gluon contribution to these quantities are
easily found from the regulated form.  It is also reasonably straightforward
to find the relation between the mass scales in a regularized vertex and
self-energy which guarantees that the Ward identity holds at the renormalized
level.

The hard-nosed, unimaginative reader will note, as is correct, that conformal
symmetry {\it per se\/} plays no direct role in our work, and our results can
be
more prosaically explained by the fact that the bare amplitudes of the
background field method in Feynman gauge are linear combinations of the two
tensor structure above.  However, it can hardly be mere coincidence that these
tensors are conformally covariant, although this curiosity is not now
understood from a general standpoint.

Further, differential regularization is particularly useful here simply because
it is a real space method in which regularized and bare amplitudes of
primitively divergent graphs agree for separated points.  The conformal
structure found for the renormalized amplitudes is independent of the
regularization method used.  In any method, it would be evident in real space
if anyone cared to look.  In momentum space, conformal transformations are
integral transformations.  Both this and scale-dependent renormalization
obscure the conformal properties.

To gain some perspective, we discuss further some non-conformal invariant
correlators such as the three-gluon vertex in conventional Feynman gauge field
theory.  Here one can write the bare amplitude as a multiple of $D^{\rm
sym}_{\mu\nu\rho}(x,y,z)$ plus a remainder which has a finite Fourier
transform.  (The coefficient of  $D^{\rm sym}_{\mu\nu\rho}$ differs from that
in the conformal invariant background field vertex.)  Thus the divergent part
of the amplitude is described by a conformal
tensor, and the reason is very simple.  All versions of the
three-gluon vertex are linearly divergent and have the full Bose permutation
symmetry.  The renormalization scale (or ultraviolet cutoff) dependence is
uniquely determined by the tensor form and discrete symmetry to be a multiple
of the tree approximation Yang--Mills vertex, namely the linear polynomial
$$V_{\mu\nu\rho}(k_1,k_2,k_3) = \delta_{\mu\nu}\left(k_1-k_2\right)_\rho
+ \delta_{\nu\rho}
\left(k_2-k_3\right)_\mu + \delta_{\rho\mu} \left( k_3-k_1\right)_\nu
\eqno(1.12)$$
in momentum space.  The regularized form of the tensor $D^{\rm
sym}_{\mu\nu\rho}(x,y,z)$ also has the same scale-dependence, and there is a
unique multiple of this tensor which fully accounts for the ultraviolet
divergence.

The one-loop vertex function with one background gluon and two quantum gluons
is a linearly divergent subgraph of the two-loop background field vertex.
Although it does not have a full Bose symmetry, the known renormalization
structure$^{11}$ of the background field formalism requires that its
renormalization scale-dependence is that of (1.12).  So it also can be written
as a multiple of $D^{\rm sym}_{\mu\nu\rho}(x,y,z)$
plus a remainder which is ultraviolet convergent.  This
representation may be useful in the study of the two-loop background field
vertex.

In Section~II, we present the background field formalism employed in our work,
and in Section~III, we discuss the computations which established the
conformal properties of the background field vertex.  In Section~IV, we
discuss the regularization of $D^{\rm sym}_{\mu\nu\rho}$ and $C^{\rm
sym}_{\mu\nu\rho}$.  The renormalized Ward identity and the mass scale
relation for the regularized vertex and self-energy are studied in
Section~V.  In Section~VI we show how the $\beta$-function is obtained from
the regulated vertex.  In Section~VII we outline a speculative scenario about
the role of conformal invariance in higher-loop calculations, and there is a
brief statement of the conclusions in
Section~VIII.  Appendix A is devoted to the study of the linear deviation from
Feynman gauge, while some results concerning mixed vertices are presented in
Appendix B.  In Appendix C we summarize our work on the conformal properties
in a general background field gauge.
\goodbreak
\bigskip
\noindent{\bf II.\quad THE BACKGROUND FIELD METHOD}
\medskip
\nobreak
We now outline the background field formalism used in our work following
Abbott$^{6}$ and the treatment in Ref.~[1] with some minor but not
insignificant changes.

Given the gauge potentials $A^a_\mu(x)$ and the structure constants $f^{abc}$
of a semi-simple Lie algebra, the Yang--Mills field strength and action are
$$\eqalign{F^a_{\mu\nu} &= \partial_\mu A^a_\nu - \partial_\nu A^a_\mu +
f^{abc} A^b_\mu A^c_\nu \cr
S[A] &= {1\over 4g^2} \int d^4x F^a_{\mu\nu} F^a_{\mu\nu} \ \ .\cr}
\eqno(2.1)$$
One now introduces the background/quantum split
$$\eqalign{A^a_\mu(x) &= B^a_\mu (x) + gb^a_\mu (x) \cr
F^a_{\mu\nu} &= B^a_{\mu\nu} + g \left( D_\mu b^a_\nu - D_\nu b^a_\mu \right)
+ g^2 f^{abc} b^b_\mu b^c_\nu \cr
D_\mu b^a_\nu &= \partial_\mu b^a_\nu + f^{abc} B^b_\mu b^c_\nu \cr}
\eqno(2.2)$$
where, unless otherwise specified, $D_\mu$ denotes a covariant derivative with
background connection.  The action $S[B+gb]$ is separately invariant under
background gauge transformations
$$\eqalign{\delta B^a_\mu &= D_\mu \theta^a\ \ ,\cr
\delta b^a_\mu &= f^{abc} b^b_\mu
\theta^c \cr}\eqno(2.3)$$
and under quantum gauge transformations
$$\eqalign{\delta B^a_\mu &= 0 \cr
\delta b^a_\mu &= D_\mu \alpha^a + gf^{abc} b^b_\mu \alpha^c \cr
&\equiv D_\mu \left( B + gb\right) \alpha^a\ \ .\cr}\eqno(2.4)$$
The gauge fixing action
$$S_{gf}[b] = {1\over 2a} \int d^4x \left( D_\mu b^a_\mu\right)^2 \eqno(2.5)$$
is invariant under background field transformations only, as is the associated
Faddeev--Popov operator
$$M[B,b] = D_\mu D_\mu (B+gb)\ \ .\eqno(2.6)$$

We define the functional
$$e^{-\Omega[B, J]} = \int [db_\mu] \det M \exp - \left\{ S[B+gb] - S[B] +
S_{gf}[b] + \int d^4x\, J^a_\mu(x) b^a_\mu\right\} \ \ .\eqno(2.7)$$
The source $J^a_\mu(x)$ is given by
$$J^a_\mu (x)={1\over g} D_\alpha B^a_{\alpha\mu}(x) + j^a_\mu(x) \ \
.\eqno(2.8)$$
The purpose of the first term is to cancel the linear ``tadpole'' in $S[B+gb] -
S[B]$, while $j^a_\mu(x)$ is the source for quantum gluons. For $j^a_\mu(x)
\equiv 0$, $\Omega [B,J]_{j=0}$ contains one-particle irreducible graphs with
external $B$ fields and internal $b$ lines beginning in one-loop order, plus
some non-1PI graphs beginning in two-loop order.  The 1PI graphs can be
systematically treated by a Legendre transform and they contribute to the
gauge invariant effective action of the theory which agrees with the
conventional effective action in an unconventional gauge.$^{5,\,6}$  The
Legendre transform is not discussed here because it is not required for the
one-loop computations which are the major part of the present work.  For our
schematic discussion of two-loop order in Section~VII, it is necessary to
note that one-loop subdiagrams with external quantum gluons are required and
these are obtained by functional differentiation of $\Omega[B,J]$ with respect
to $j^a_\mu$ followed by amputation on external $b$-lines.

The most important property of the background field formalism for our purpose
is its invariance under background gauge transformations (2.3).  A gauge
transformation of $B^a_\mu$ is compensated, except in the source term
involving $j_\mu$, by a gauge rotation of the integration variable $b_\mu$,
leading to the functional Ward identity
$$D_\mu {\delta \Omega [B,J]\over \delta B^a_\mu (x)} = - f^{abc} j^b_\mu (x)
{\delta\over \delta j^c_\mu (x)} \Omega[B,J] \ \ .\eqno(2.9)$$
When there are no external quantum gluons the right side vanishes.
As a special case of (2.9) one finds that the background three-gluon vertex
and self-energies are related by
$$\eqalign{
{\partial\over\partial z_\rho} \ {\delta^3\Omega[B,J]_{j=0} \over \delta
B^a_\mu(x) \delta B^b_\nu(y) \delta B^c_\rho (z)} = &- f^{cad} \delta(z-x)
{\delta^2\Omega[B,J]_{j=0}\over \delta B^d_\mu (x) \delta B^b_\nu(y)} \cr
&- f^{cbd}\delta(z-y) {\delta^2\Omega [B,J]_{j=0} \over \delta B^a_\mu (x)
\delta B^d_\nu (y)} \cr}\eqno(2.10)$$
which is exactly of the same form as the identity (1.4) satisfied by current
correlation functions.

To implement perturbation theory, one needs the explicit form of the integrand
of (2.7)
$$\eqalign{
S[B+gb] &- S[B] + S_{gf} [B]
+ {1\over g} \int d^4x\, D_\alpha B^a_{\alpha\mu} b^a_\mu \cr
&=
\int d^4x\, \left[ {1\over 2} D_\mu b^a_\nu D_\mu b^a_\nu + f^{abc}
B^a_{\mu\nu} b^b_\mu b^c_\nu + {1\over 2} \left( {1\over a} - 1\right) \left(
D\cdot b\right)^2 + {\cal L}_q \right]\cr} \eqno(2.11)$$
where
$${\cal L}_q = gf^{abc} \left( D_\mu b^a_\nu\right) b^b_\mu
b^c_\nu + {g^2\over 4} f^{abc} f^{ade} b^b_\mu b^c_\nu b^d_\mu b^e_\nu
\eqno(2.12)$$
describes quantum gluon vertices which
are required in background field calculations beyond one-loop.  As
in [1] we rewrite the integrand of (2.11) in terms of $b_\mu$ kinetic terms
and mixed $b$-$B$ interaction terms
$$\eqalign{
{\cal L}_0 &= {1\over 2} \partial_\mu b^a_\nu \partial_\mu b^a_\nu + {1\over
2} \left( {1\over a} - 1\right) \partial_\mu b^a_\mu \partial_\nu b^a_\nu \cr
{\cal L}_1 &= f^{abc} B^a_\mu b^b_\nu \partial_\mu b^c_\nu \cr
{\cal L}_2 &= {1\over 2} f^{abc} f^{ade} B^b_\mu B^d_\mu b^c_\nu b^e_\nu \cr
{\cal L}_3 &= f^{abc} B^a_{\mu\nu} b^b_\mu b^c_\nu \cr
{\cal L}_4 &= \left( {1\over a} - 1\right) \left[ f^{abc} B^b_\mu b^c_\mu
\partial_\nu b^a_\nu + {1\over 2} f^{abc} f^{ade} B^b_\mu b^c_\mu B^d_\nu
b^e_\nu\right] \ \ .\cr} \eqno(2.13)$$
The quantum field propagator is
$$\ll b^a_\mu (x) b^b_\nu(y)\rr = {\delta^{ab}\over 4\pi^2} \left[
{\delta_{\mu\nu}\over (x-y)^2} + {a-1\over 4} \partial_\mu \partial_\nu \ln
(x-y)^2\right]\ \ .\eqno(2.14)$$
One sees that both propagator and interaction terms simplify in the Feynman
gauge $(a=1)$ which was the initial motivation for this internal gauge choice.

For computational purposes, one represents the Faddeev--Popov determinant as a
functional integral over anti-commuting ghosts $c^a(x)$, $\bar{c}^a(x)$ with
action
$$\eqalign{S_{\rm gh}[c,\bar{c}] &= \int d^4x \left[ D_\mu \bar{c}^a D_\mu c^a
+ {\cal L}'_g \right]\cr
&= \int d^4x \left[{\cal L}^g_0 + {\cal L}^g_i+ {\cal L}^{'g} \right] \cr}
\eqno(2.15)$$
and
$$\eqalign{
{\cal L}^g_0 &= \partial_\mu \bar{c}^a \partial_\mu c^a\cr
{\cal L}^g_i &= f^{abc} B^a_\mu \bar{c}^b \upleftrightarrow{\partial}_\mu c^c
+ f^{abc} f^{ade} B^b_\mu B^d_\mu \bar{c}^c c^e \cr
{\cal L}^{'g} &= gf^{abc} D_\mu \bar{c}^a b^b_\mu c^c\ \ .\cr} \eqno(2.16)$$
Note that $\upleftrightarrow{\partial}_\mu$ is an anti-symmetric derivative,
and that the ghost propagator is
$$\ll c^a(x) \bar{c}^b(y) \rr = {1\over 4\pi^2}\ {\delta^{ab}\over (x-y)^2} \
\ .\eqno(2.17)$$

The only new feature of this treatment, compared with Refs. [1] and [6] is
that the tadpole terms were simply neglected previously, because they do not
contribute to 1PI diagrams.  Now the tadpole is cancelled explicitly, through
the form (2.8) of the source.  This makes a difference only for some of the
functional identities used in Section~III and Appendix~A.

There are more recent versions of the background field formalism$^{12}$
 which are
more general than that used here.  When applied to non-Abelian gauge theories,
they agree with the present version to one-loop order, if the Landau gauge,
rather than the Feynman gauge, is used for internal gluons.  In two-loop
order, there are other differences.
Since the three-gluon vertex in the Landau gauge is not conformal
invariant it does not seem that these formalisms are useful for
further investigation of the conformal property.
\goodbreak
\bigskip
\noindent{\bf III.\quad BARE AMPLITUDES AND CONFORMAL TENSORS}
\medskip
\nobreak
One-particle irreducible (1PI) diagrams
contributing to the one-loop correction to the three-gluon
vertex can be classified in three groups, each of which provides a different
conformally invariant contribution to the effective action.  Graphs with (1)
ghost loops, (2) fermion loops,and (3) gluon loops are separately conformally
invariant,namely they are linear combinations of the conformal tensors $C^{\rm
sym}_{\mu\nu\rho}(x,y,z)$ and $D^{\rm sym}_{\mu\nu\rho}(x,y,z)$ of (1.8 --
1.10), with different coefficients.  The conformal property of the ghost and
fermion diagrams are clearly associated with invariant Lagrangians, but that
of the gluon graphs presently lacks a simple explanation.
In this section we describe in some
detail the calculations that led us to recognize the conformal property, which
is a regularization-independent result requiring, however, a real space
approach.

{}From (2.15) one sees that the part of the ghost Lagrangian required in
one-loop calculations is simply $D_\mu \bar{c}^a D_\mu c^a$ which coincides
with that of a minimally coupled scalar field in the adjoint representation.
This is conformal invariant, as one can see, for example, by combining
(1.11) for $B^a_\mu(x)$ with
$$c^{'a}(x') = x^2  c^a(x) \eqno(3.1)$$
for the ghost (and antighost).  Thus the ghost contribution to the three-gluon
vertex will be conformal invariant at one-loop order. At the computational
level, it is the antisymmetric derivative $\upleftrightarrow{\partial}_\mu$ in
(2.16) that is crucial for the conformal property.  The ghost interaction
term ${\cal L}'{}^g$ of (2.15) is not conformal invariant.

To see which linear combination of the conformal tensors $C^{\rm
sym}_{\mu\nu\rho}(x,y,z)$ and $D^{\rm sym}_{\mu\nu\rho}(x,y,z)$ describes the
ghost contribution we examine the two 1PI diagrams of Fig.~1.  Graph (1.a)
vanishes because the Wick contractions give an algebraic factor of the type
$f^{abc} B^a_\mu(x) B^b_\mu(x)~=~0$.  Graph (1.b) instead gives the following
contribution to $-\Omega_3[B]$, {\it i.e.\/} the part of the effective action
cubic in $B$,
$$\eqalign{
\hbox{(1.b)} &= -{1\over 3}
 \  {C\over \left(4\pi^2\right)^3} f^{abc} \int d^4x\, d^4y\, d^4z\, B^a_\mu
(x)
B^b_\nu(y) B^c_\rho(z)\cr
&\times \left[ T_{\mu\rho\nu}(x-z,y-z) - {3\over 2}
V_{\mu\nu\rho} (x-z, y-z) \right] \cr} \eqno(3.2)$$
where we have introduced the tensors
$$\eqalignno{ T_{\mu\nu\rho} (x-z,y-z) &= \partial^x_\mu {1\over (x-z)^2}
\partial^y_\nu {1\over (y-z)^2} \partial^x_\rho {1\over (x-y)^2} \ \
,&(3.3\hbox{a}) \cr
V_{\mu\nu\rho} (x-z, y-z) &= \partial^x_\mu \partial^x_\nu {1\over (x-y)^2} \
{1\over (y-z)^2} \upleftrightarrow{\partial}^z_\rho {1\over (x-z)^2}\ \ ,
&(3.3\hbox{b}) \cr}$$
$C$ is the Casimir operator in the adjoint representation ($C=N$ for $SU(N)$)
and we have used $f^{ade} f^{bef} f^{cfd} = {C\over 2} f^{abc}$.  For the
contribution to the three-point function from (3.2) we get
$$\eqalign{
{\delta^3\hbox{(1.b)}\over \delta B^a_\mu(x) \delta B^b_\nu(y) \delta
B^c_\rho(z)} &=-{C\,f^{abc}\over \left( 4\pi^2\right)^3} \biggl[
T_{\mu\rho\nu} (x-z, y-z) + T_{\rho\nu\mu} (x-z, y-z)\cr
&\qquad  - {3\over 2} V^{\rm
sym}_{\mu\nu\rho} (x-z, y-z)\biggr]\ \ ,\cr}\eqno(3.4)$$
where $V^{\rm sym}_{\mu\nu\rho}$ is constructed as in (1.10) by adding cyclic
permutations.  By manipulating the derivatives it is possible to
show that (3.4) can be rewritten in terms of the conformal tensors of
(2.10) according to
$${\delta^3 \hbox{(1.b)}\over \delta B^a_\mu(x) \delta B^b_\nu(y) \delta
B^c_\rho(z)} =  -{C\over \left( 4\pi^2\right)^3}\ {f^{abc}\over3} \left[
{1\over
2} C^{\rm sym}_{\mu\nu\rho}(x,y,z) + 4D^{\rm sym}_{\mu\nu\rho}(x,y,z)\right] \
\ .\eqno(3.5)$$

It is worth mentioning that if the analogous quantity is computed in the
usual quantum field theoretical approach, instead of the background field
method, one would get a non-conformally invariant amplitude proportional to the
quantity $T_{\mu\rho\nu}(x-z,y-z) + T_{\rho\nu\mu}(x-z,y-z)$.  This is to be
expected since the ghost action for conventional Lorentz gauge fixing is not
conformally invariant.

Let us now analyze the fermion loop contributions.
The Euclidean fermion action is (we assume only one flavor)
$$S^F = i \int d^4x\,\bar{\psi} \gamma_\mu \left( \partial_\mu + A_\mu\right)
\psi = \int d^4x \left( {\cal L}^F_0 + {\cal L}^F_i\right) \eqno(3.6)$$
where $A_\mu = - i A^a_\mu T^a$, $T^a$ are the Hermitian gauge group
generators and $\left\{ \gamma_\mu,\gamma_\nu\right\} = 2\delta_{\mu\nu}$ are
Euclidean Hermitian Dirac matrices.  The action (3.6) is conformally invariant
and the conformal properties of the fermion loop triangle were already studied
in the Abelian case.$^{10}$
In Yang--Mills theories there are two diagrams contributing to the
one-loop three-gluon vertex (Fig.~2).  The fermion propagator in real space
reads
$$\ll \psi^i(x) \bar{\psi}^j(y) \rr = - {i\over 4\pi^4} \delta^{ij}
\gamma^\mu \partial^x_\mu {1\over (x-y)^2} \eqno(3.7)$$
where $i,j=1,\ldots,N$ for $SU(N)$
are the representation indices.  Both diagrams
are obtained from the Wick contractions in $\ll {\cal L}^F_i (x) {\cal
L}^F_i(y) {\cal L}^F_i (z) \rr$ which gives the contribution to $\Omega_3[B]$
$$\eqalign{\hbox{(2.a)} + \hbox{(2.b)} &=-{1\over 3} \ {i\over \left(
4\pi^2\right)^3} \int d^4x\, d^4y\, d^4z\, B^a_\mu(x) B^b_\nu(y) B^c_\rho(z)
\cr
&\times \left[ \Tr \left\{ T^b T^a T^c\right\} \cdot \tr \left\{
\gamma_\lambda \gamma_\mu \gamma_\sigma \gamma_\rho \gamma_\tau
\gamma_\nu\right\} T_{\sigma\tau\lambda} (x-z, y-z) \right] \ \ .\cr}
\eqno(3.8)$$
The trace of three generators is
$$\Tr \left\{ T^a T^b T^c\right\} = {1\over 4} \left( d^{abc} + i
f^{abc}\right) \eqno(3.9)$$
and in (3.8) the terms containing the symmetric part $d^{abc}$ vanish by
symmetry properties of the trace on the $\gamma$ matrices.  Using the trace
one finds, after a simple calculation, the explicit conformal invariant
expression for the fermion loop contribution to the vertex
$${\delta\left[ (\hbox{2.a)} + \hbox{(2.b)} \right]\over \delta B^a_\mu(x)
\delta B^b_\nu (y) \delta B^c_\rho (z)} = {4\over 3} \  {f^{abc}\over \left(
4\pi^2\right)^3} \left[ - {1\over 2} C^{\rm sym}_{\mu\nu\rho} (x,y,z) +
2D^{\rm sym}_{\mu\nu\rho} (x,y,z)\right] \ \ .\eqno(3.10)$$

The conformal invariance of the one-loop diagrams containing gluon loops is
much more surprising, since the gauge fixing term (2.5) breaks conformal
invariance and, as discussed in the introduction, the gluon propagator does
not transform properly.
The 1PI gluon diagrams are shown in Fig.~3.  It turns
out that diagram (3.e) is separately conformal invariant in Feynman  gauge
($a=1$) because the antisymmetric $\upleftrightarrow{\partial}_\mu$ in ${\cal
L}_1$ of (2.13) does ``match'' nicely with the effective ``scalar-like''
gluon propagator of (2.14).  Indeed the amplitude of (3.e) is simply
$-2$ times that of the ghost loop (3.5) and thus embodies the expected ghost
cancellation of two of the four degrees of freedom of $b_\mu$.

The remaining
diagrams of Fig.~3 are not separately conformal invariant but their sum is.
We describe the calculations as follows.  The graphs (3.a,b,c,f) vanish
because symmetric tensors in the group indices are contracted with $f^{abc}$.
For the graph (3.h) we have the interesting property that, after partial
integration of all derivatives, the terms in which Dirac $\delta$ occur, {\it
i.e.\/} terms in which two background fields are at the same point cancel
exactly with the seagull diagrams (3.d).  The same type of local terms,
instead cancel among themselves in the graph (3.g).  Adding up all these
contributions, after a lengthy but straightforward calculation, it is possible
to express also the gluon loop contributions to the three point function in
terms of the conformal tensors
$${\delta^3 \left( \hbox{(3.d)} + \hbox{(3.e)} + \hbox{(3.g)} + \hbox{(3.h)}
\right]\over \delta B^a_\mu(x) \delta B^b_\nu (y) \delta B^c_\rho(z)} =
{C\over \left( 4\pi^2\right)^3}\ {f^{abc}\over 3} \left[ 7 C^{\rm
sym}_{\mu\nu\rho} (x,y,z) - 40 D^{\rm sym}_{\mu\nu\rho} (x,y,z)\right] \ \
.\eqno(3.11)$$

In conclusion, the one-loop correction to the three-gluon vertex, computed in
the background field framework in Feynman gauge, is conformal invariant.  The
sum of all contributions of the three groups of graphs for the three point
function with $N_f$ fermions is
$$\eqalign{ {\delta^3\Omega [B,J]\over \delta B^a_\mu (x) \delta B^b_\nu(y)
\delta B^c_\rho(z)} = {f^{abc}\over 3\left( 4\pi^2\right)^3} \biggl[ &-\left(
{13\over 2} C - 2N_f\right) C^{\rm sym}_{\mu\nu\rho} (x,y,z) \cr
&+ \left( 44C - 8N_f\right) D^{\rm sym}_{\mu\nu\rho} (x,y,z) \biggr]\ \
.\cr} \eqno(3.12)$$
The coefficients of the divergent tensor in (3.12) are exactly those
necessary to satisfy the Ward identity (2.10) and, consequently they are
directly related to the well-known $-{11\over 3} C+{2\over 3} N_f$ of the
Yang--Mills $\beta$ function.  The details of these questions are discussed in
Sections~V and VI.

The unexpected conformal property of the gluon graphs requires explanation,
and we have attempted to explain it via a Slavnov--Taylor-like identity which
describes the conformal variation of the generating functional
$\Omega[B,J]_{j=0}$.  The infinitesimal form of the conformal transformation
(1.3) is easily described$^{13}$
in terms of conformal Killing vectors defined as
$$x_\mu \longrightarrow x_\mu + v_\mu (x,\epsilon) \equiv x_\mu + \epsilon_\mu
x^2 - 2x_\mu \epsilon\cdot x\ \ .\eqno(3.13)$$
These vectors satisfy
$$\eqalign{\partial_\mu v_\nu &+ \partial_\nu v_\mu - {1\over 2}
\delta_{\mu\nu}
\partial\cdot v = 0 \cr
\partial\cdot v &= - 8\epsilon\cdot x\cr
\square v_\mu &= 4\epsilon_\mu \ \ .\cr}\eqno(3.14)$$
The standard conformal transformations of vector and scalar fields are
$$\eqalign{\delta'_\epsilon A_\mu &= v_\nu \partial_\nu A_\mu + A_\nu
\partial_\mu v_\nu \cr
\delta'_\epsilon \varphi &= v_\nu \partial_\nu \varphi - 2 \epsilon\cdot x
\varphi\ \ .\cr}\eqno(3.15)$$
 Because of the background gauge invariance of our formalism, it is more
convenient to add the field-dependent gauge transformation with parameter
$\theta^a=v_\rho B^a_\rho$ and use the gauge covariant conformal
variations$^{14}$
$$\eqalign{\delta_\epsilon B^a_\mu &=- v_\nu B^a_{\nu\mu}\cr
\delta_\epsilon b^a_\mu &= v_\nu D_\nu b^a_\mu + b_\nu \partial_\mu v_\nu\cr
\delta_\epsilon c^a &= v_\nu D_\nu c^a - 2\epsilon\cdot x \, c^a \ \ .\cr}
\eqno(3.16)$$

It is not difficult to see that $S[B+bg]$ and $S[B]$ are conformal invariant,
and that
$$\delta_\epsilon D\cdot b^a = \left( v \cdot D + {1\over 2} \partial\cdot
v\right) D\cdot b^a + 4\epsilon\cdot b\eqno(3.17)$$
leading to the simple variation of the gauge fixing term (2.5)
$$\delta_\epsilon S_{gf} [b] = {4\over a} \int d^4x\,\epsilon\cdot b^a D\cdot
b^a \ \ .\eqno(3.18)$$
Thus the effect of a conformal transformation of $D^a_\mu$ and $b^a_\mu$ is to
change the gauge fixing $D\cdot b^a$ as follows
$$D\cdot b^a\longrightarrow D\cdot b^a + 4\epsilon\cdot b^a \ \ .\eqno(3.19)$$

We now describe qualitatively the effect of a conformal transformation on
$\Omega[B,J]$ in (2.7).  Ignoring ghost variations for the moment, we combine
transformations of the background $B^a_\mu$ and source
$$\delta_\epsilon
J^a_\mu = v_\nu D_\nu J^a_\mu + J^a_\nu \partial_\mu v_\nu + {1\over
2} \partial\cdot v J_\mu\ \ \eqno(3.20)$$
and make the analogous change in the integration variable $b_\mu$.  The net
effect in (2.7) is the change (3.18) of the gauge fixing term.  We then make
the further quantum gauge transformation
$$b^a_\mu = b^{'a}_\mu - D_\mu [B+b] \theta^a\ \ ,\qquad \theta^a = 4M^{-1}
\,\epsilon \cdot b^a\eqno(3.21)$$
to restore the original gauge and transfer the change to the source term.  The
process just described leads to the functional identity (for connected graphs)
$$\eqalign{{1\over 4} \delta_\epsilon \Omega[B,J] &= {1\over a} \ll \int
d^4x\,\epsilon\cdot b^a \,D\cdot b^a\rr\cr
&= \ll \int d^4x\, J^a (x)\cdot D [B+b] \int d^4y\, M^{-1} (x,y)^{ab}
\epsilon\cdot b^b (y)\right.\cr
&\quad + \left. \int d^4x \, \epsilon \cdot D[B+b] M^{-1}(x,y=x)^{aa'=a}
\rr\cr
&= \ll \int d^4x\, J^a(x) \cdot D[B+b] c^a(x) \int d^4y\, \bar{c}^b (y)
\epsilon\cdot b^b (y) \right.\cr
&\quad + \left. \int d^4x\,\epsilon\cdot D[B+b] c^a (x) \bar{c}^a(x) \rr \ \
.\cr}\eqno(3.22)$$
The first term on the second line is the induced change in the source term,
and the next term is the formal contribution of the change of the integration
measure due to (3.21).

What about ghosts?  We already know that the ghost action is conformal
invariant if $b^a_\mu=0$, and it is easy to confirm this using (3.16).  There
are no other ghost variation effects at the one-loop level, and the functional
identity (3.22) correctly expresses the conformal variation of $\Omega[B,J]$
and the bare one-loop amplitudes it generates.  An alternative way to pass
from the first to the last line of (3.22) is to apply $\int
d^4x\,\epsilon\cdot \left( \delta/\delta J^a_\mu(x)\right)$ to the functional
Slavnov--Taylor identity given in (A.2).  At higher order there are
additional ghost effects which we have not treated.  Thus (3.22) may well be a
one-loop-only result,  but that is all we need.

To see if (3.22) sheds any light on our background field results, we set
$j^a_\mu=0$.  Using (2.8) and its covariant conservation
property, we rewrite (3.22) as
$$\eqalign{{1\over 4} \delta_\epsilon\Omega [B,J]_{j=0} &= \ll f^{abc} \int
d^4x\, D_\nu B^a_{\nu\mu} (x) b^b_\mu (x) c^c(x) \int d^4y\, \bar{c}^d(y)
\epsilon\cdot b^d(y)\right.\cr
&\quad \left. + \int d^4x\left( \epsilon\cdot D [B+b] c^a(x) \right) \bar{c}^a
(x) \rr \ \ .\cr}\eqno(3.23)$$
To account for the experimental results of this paper, the third variational
derivative of (3.23) with respect to $B^a_\mu$ must vanish in the Feynman
gauge but not for general values of $a$.  One sees no direct reason for this
in (3.23), other than through a detailed computation of the contributing
diagrams which would be a tedious job.  As a check against possible error in
(3.23), we did study the diagrams which contribute to the second variational
derivative.  Here it is not difficult to show that contributions to the first
and second term in (3.23) vanish separately.  This confirms that the
background field self-energy is conformal invariant, and is consistent with
the fact that the form (1.6) is obtained from direct calculation.

Three-gluon vertices with one or more external quantum gluons are studied in
the Appendices.  It can be argued very simply, using Ward identities, that
they cannot be conformal invariant.  It is implicit in the discussions of
(1.4) -- (1.7) of the Introduction, Section~I, and confirmed in Sections~IV
and V below, that a linear combination of conformal tensors $D^{\rm
sym}_{\mu\nu\rho}$ and $C^{\rm sym}_{\mu\nu\rho}$ satisfies a simple Ward
identity, specifically that the divergence $\partial/\partial z_\rho$ produces
a sum similar to (1.4) of two gauge-invariant self-energies of the form (1.6).
 However, the divergence of the vertex function for three external quantum
gluons involves the more complicated mathematical structure of the
Slavnov--Taylor identity (A.3).  Thus the three-gluon vertex of conventional
field theory cannot be conformal invariant.

Mixed vertices with both background and quantum external gluons, satisfy
simple Ward identities when the divergence is taken in the background field.
This follows from (2.6) and the fact that mixed and quantum self-energies both
take the gauge invariant form (1.6).  However, the divergence on a quantum
vertex is again more complicated, as one can see from (A.3).  This is not quite
enough to conclude that mixed vertices are not conformal invariant, because
the mixed vertices have the reduced Bose symmetry of a single conformal tensor
$D_{\mu\nu\rho}(x,y,z)$ and this tensor also has the curious property that the
$\partial/\partial z_\rho$ divergence is that of a simple Ward identity,
while the $\partial/\partial x_\mu$ and $\partial/\partial y_\nu$ divergences
are more complicated.

The argument that the mixed vertices are not conformal invariant can be
completed in several ways, and we choose an argument which gives an additional
piece of information.  The renormalization properties of the background field
formalism have been studied by Kluberg--Stern and Zuber.$^{11}$  They find
that the counterterm for the overall divergence of a three-gluon vertex with
any combination of external background and quantum lines takes the form of the
bare Yang--Mills vertex, that is the cubic term of (2.1).  This is confirmed
by our one-loop calculations, and it has the implication that the part of the
one-loop amplitude which requires regularization can be written as a numerical
multiple of $D^{\rm sym}_{\mu\nu\rho}$.  The remainder is ultraviolet
finite.  For mixed vertices, this means that the remainder cannot simply be a
non-fully symmetric combination of the linearly independent tensors
$D_{\mu\nu\rho}$ and cyclic permutations, because any such combination requires
regularization.  Mixed vertices are therefore not conformal invariant,
although we hope that the fact that their divergent part is a multiple of
$D^{\rm sym}_{\mu\nu\rho}$ will facilitate study of the Slavnov--Taylor
identities and help with two-loop calculations in the background field method.

It is relevant to ask whether other one-loop vertex functions of gauge
theories can have the conformal properties found here.  The example (1.1) of
the electron vertex function in quantum electrodynamics shows that this is not
the case, and we discuss this briefly here.  The inversion property of a
fermion field is
$$\psi'(x') = x^2\gamma_5 \gamma\cdot x\,\psi(x)\ \ .\eqno(3.24)$$
Using this and (1.11), it is not difficult to see that the amplitude
$$\tilde V_\lambda (x,y,z) = - 2\gamma_a \gamma_\lambda \gamma_b V_{ab}
(x-z,y-z)\eqno(3.25)$$
transforms properly under conformal transformation, but (1.1) does not.
Indeed $\tilde V_\lambda$ is the one-loop electromagnetic vertex in a
Lagrangian in which a fermion is coupled via $\bar{\psi}\psi\phi$ to a
massless scalar field, so conformal invariance is expected!  The blame for the
non-invariance in the case of quantum electrodynamics rests squarely on the
shoulder of the gauge fixing procedure, which affects the virtual photon
propagator.  One can easily see from (1.1) that the difference
$V_\lambda-\tilde V_\lambda$ is a total derivative and therefore ultraviolet
finite.  Thus we find again that the part of a non-conformal vertex that
requires regularization is conformal for non-coincident points.
\goodbreak
\bigskip
\noindent{\bf IV.\quad REGULARIZATION OF CONFORMAL COVARIANT TENSORS}
\medskip
\nobreak
In the previous section the bare, primitive divergent Feynman amplitudes for
the three-gluon vertex were expressed in terms of the conformal tensors
$D^{\rm sym}_{\mu\nu\rho}$ and $C^{\rm sym}_{\mu\nu\rho}$.  The regularization
problem  for the physical amplitudes is therefore solved by regulating these
tensors, and regularization is required because the short-distance
singularities make the Fourier transforms diverge.  Indeed, each tensor
$D_{\mu\nu\rho}$ and $C_{\mu\nu\rho}$ of (1.8) and (1.9) corresponds to a
formally linear divergent loop integral in momentum space.  However,
the combination $C^{\rm sym}_{\mu\nu\rho}$ of (1.10) is
ultraviolet finite.  Thus the various
contributions to the three-gluon vertex involve a universal divergent tensor
$D^{\rm sym}_{\mu\nu\rho}$.  In this section we present two distinct
regularized expressions for $D^{\rm sym}_{\mu\nu\rho}$ using the method of
differential regularization.  We also discuss the properties of $C^{\rm
sym}_{\mu\nu\rho}$ which, although finite, requires regularization to make its
Fourier transform unambiguous.

In the first approach to regularize the conformal tensor
$D_{\mu\nu\rho}(x,y,z)$, it is convenient to write it in terms of the tensors
$T_{\mu\nu\rho}(x-z,y-z)$ and $V_{\mu\nu\rho}(x-z,y-z)$ introduced in
(3.3).  Note that $T_{\mu\nu\rho}$ has been regulated already in
Ref.~[1] and $V_{\mu\nu\rho}$ may be easily regulated following  the same
basic  ideas.  A straightforward calculation shows that
$$D_{\mu\nu\rho}(x,y,z) = {1\over 4} \delta_{\mu\nu} \left(
T_{\lambda\rho\lambda} + T_{\rho\lambda\lambda} - T_{\lambda\lambda\rho}
\right) - {1\over 2} V_{\mu\nu\rho} (x-z,y-z) \eqno(4.1)$$
where $T_{\alpha\beta\gamma}\equiv T_{\alpha\beta\gamma}(x-z,y-z)$ here and
in the following.

Now, we briefly summarize the regularization of $T_{\mu\nu\rho}$ given in
Ref.~[1].  The procedure maintains explicitly the $x\leftrightarrow y$,
$\mu\leftrightarrow\nu$ antisymmetry of the tensor and consists of moving the
derivatives to the left in order to have a piece with two total derivatives,
which has a finite Fourier transform, and some remaining terms whose singular
parts lie only in the trace.  Thus, $T_{\mu\nu\rho}$ may be written as
$$T_{\mu\nu\rho}(x,y) = F_{\mu\nu\rho} (x,y) + S_{\mu\nu\rho} (x,y)\ \
,\eqno(4.2)$$
where we have set $z=0$ for simplicity.  $F_{\mu\nu\rho}(x,y)$, which has a
well-defined Fourier transform, is
$$\eqalign{F_{\mu\nu\rho} (x,y) &= \partial^x_\mu \partial^y_\nu \left[ {1\over
x^2 y^2} \partial^x_\rho {1\over (x-y)^2}\right]
+ \partial^x_\mu \left[ {1\over x^2 y^2} \left( \partial_\nu \partial_\rho -
{1\over 4} \delta_{\nu\rho} \square\right) {1\over (x-y)^2}\right] \cr
&- \partial^y_\nu \left[ {1\over x^2 y^2} \left( \partial_\mu \partial_\rho -
{1\over 4} \delta_{\mu\rho} \square \right) {1\over (x-y)^2} \right] \cr
&- {1\over x^2 y^2} \left[ \partial^x_\mu \partial^x_\nu \partial^x_\rho -
{1\over 6} \left( \delta_{\mu\nu} \partial^x_\rho + \delta_{\mu\rho}
\partial^x_\nu + \delta_{\nu\rho} \partial^x_\mu\right) \square \right] {1\over
(x-y)^2} \ \ ,\cr}\eqno(4.3)$$
while $S_{\mu\nu\rho}(x,y)$ contains the trace terms and thus derivatives of
$\delta(x-y)$ times the factor $1/x^4$, which can be regularized in the usual
way, {\it i.e.\/}, by using the identity
$${1\over x^4} = - {1\over 4} \square {\ln M^2 x^2\over x^2}\ \ ,\eqno(4.4)$$
yielding
$$S_{\mu\nu\rho}(x,y) = - {\pi^2\over 12} \left[ \delta_{\mu\nu} \left(
\partial^x_\rho - \partial^y_\rho\right) + \delta_{\mu\rho} \left(
\partial^x_\nu + 2 \partial^y_\nu\right) - \delta_{\nu\rho} \left(
2\partial^x_\mu + \partial^y_\mu\right) \right] \delta(x-y) \square {\ln M^2
x^2\over x^2} \ \ .\eqno(4.5)$$

With the same procedure, it is easy to see that also the singular part of
$ V_{\mu\nu\rho}$ lies only in the trace, and, when it is regularized by means
of
(4.4), we get
$$\eqalign{
&V_{\mu\nu\rho}(x,y) = \left( \partial^y_\rho - \partial^x_\rho\right) \left[
{1\over x^2 y^2} \left( \partial_\mu \partial_\nu - {1\over 4} \delta_{\mu\nu}
\square \right) {1\over (x-y)^2} \right] \cr
&+ {2\over x^2 y^2} \left[ \partial^x_\mu \partial^x_\nu \partial^x_\rho -
{1\over 6} \left( \delta_{\mu\nu} \partial^x_\rho + \delta_{\mu\rho}
\partial^x_\nu + \delta_{\nu\rho} \partial^x_\mu\right) \square \right]
{1\over (x-y)^2} \cr
&+ {\pi^2\over 12} \left[ - \delta_{\mu\nu} \left( \partial^x_\rho -
\partial^y_\rho\right) + 2\delta_{\mu\rho} \left( \partial^x_\nu -
\partial^y_\nu\right) + 2 \delta_{\nu\rho} \left( \partial^x_\mu -
\partial^y_\mu\right) \right] \delta(x-y) \square {\ln M^2 x^2 \over x^2}
\ \ .\cr}\eqno(4.6)$$
Notice that the second and third rank tensor traces in (4.3) and (4.6) are
independent and a regularization with several independent mass scale
parameters$^1$ $M_i$ is consistent with the $x\leftrightarrow y$,
$\mu\leftrightarrow \nu$ antisymmetry, and could have been used.  However, in
the cyclically symmetric combinations $C^{\rm sym}_{\mu\nu\rho}$ and $D^{\rm
sym}_{\mu\nu\rho}$, which are the final objects of interest for the three-gluon
vertex, the mass ratios $\ln M_i/M_j$ appear as coefficients of the bare
Yang--Mills vertex ((1.12) in $p$-space).  Thus the mass ratio ambiguity
simply corresponds to a finite choice of regularization scheme in differential
regularization, and we have chosen the simplest scheme in which only a single
scale $M$ appears from the beginning.

Substituting the regulated forms of $T_{\mu\nu\rho}$ and $V_{\mu\nu\rho}$ in
(4.1), one finds the following regulated expression for the tensor
$D_{\mu\nu\rho}(x,y,z)$:
$$\eqalign{&D_{\mu\nu\rho} (x,y,z) = {1\over 2} \left(\partial^x_\rho -
\partial^y_\rho\right) \left[ {1\over (x-z)^2 (y-z)^2} \left(\partial_\mu
\partial_\nu - {1\over 4} \delta_{\mu\nu} \square \right) {1\over
(x-y)^2}\right] \cr
&- {1\over (x-z)^2 (y-z)^2} \left[ \partial^x_\mu \partial^x_\nu
\partial^x_\rho - {1\over 6} \left( \delta_{\mu\nu} \partial^x_\rho +
\delta_{\mu\rho} \partial^x_\nu + \delta_{\nu\rho} \partial^x_\mu \right)
\square \right] {1\over (x-y)^2} \cr
&+ {1\over 4} \delta_{\mu\nu} \left( \partial^x_\lambda \partial^y_\rho +
\partial^x_\rho \partial^y_\lambda - \delta_{\lambda\rho} \partial^x_\sigma
\partial^y_\sigma\right) \left[ {1\over (x-z)^2 (y-z)^2} \partial^x_\lambda
{1\over (x-y)^2}\right] \cr
&- {\pi^2\over 12} \left\{ - 2 \delta_{\mu\nu} \left( \partial^x_\rho -
\partial^y_\rho\right) + \delta_{\mu\rho} \left( \partial^x_\nu -
\partial^y_\nu\right) + \delta_{\nu\rho} \left( \partial^x_\mu -
\partial^y_\mu\right) \right\}\left[
 \delta(x-y) \square {\ln M^2(x-z)^2\over (x-z)^2}
\right]\ \ .\cr}\eqno(4.7)$$
{}From this equation, the regulated form of $D^{\rm sym}_{\mu\nu\rho}(x,y,z)$
may be easily obtained by adding the cyclic permutations
$D_{\nu\rho\mu}(y,z,x)$ and $D_{\rho\mu\nu}(z,x,y)$.

The Fourier transform of (4.7), computed using formal integration by parts of
the total derivatives, involves a sum of essentially conventional Feynman loop
integrals which are convergent because they contain a combination of external
momentum factors and traceless tensors.  Standard methods, for example,
combining denominators using Feynman parameters, can be used to evaluate the
loop integrals.

We now discuss an alternative differential regularization of the singular
vertex amplitude $D_{\mu\nu\rho}(x,y,z)$ of (1.8).  We first put
$D_{\mu\nu\rho}$ into a form which suggests a simple regularization by
writing out (1.8) as
$$D_{\mu\nu\rho} (x,y,z) = {1\over 3} \left[ \left( \delta_{\mu\nu} \square -
\partial_\mu \partial_\nu\right) {1\over (x-y)^4} \right] \left(
{(x-z)_\rho\over (x-z)^2} - {(y-z)_\rho\over (y-z)^2}\right) {(x-y)^2\over
(y-z)^2 (x-z)^2} \ \ .\eqno(4.8)$$
This form of the bare amplitude contains the product of a factor more singular
than in the original form times a vanishing factor, and it is analogous to
the form used to regulate the primitive non-planar three-loop graph for the
four-point function of $\phi^4$-theory.$^{1}$  We then regulate the
singular factor using (4.4) and obtain
$$\eqalign{
D_{\mu\nu\rho} (x,y,z) &= {1\over 24} \left[ \left( \delta_{\mu\nu} \square -
\partial_\mu \partial_\nu\right)\square \left( {\ln M^2(x-y)^2\over (x-y)^2}
\right)\right]\cr
&\qquad \times (x-y)^2 \left( {\partial\over\partial x_\rho} - {\partial\over
\partial y_\rho}\right) {1\over (x-z)^2(y-z)^2}\ \ .\cr}\eqno(4.9)$$

We will demonstrate that this expression gives a satisfactory regularization
by showing that it has a well-defined Fourier transform when the formal
partial integration rule of differential regularization is used.  This
regularized form is considerably simpler than the previous (4.7), although it
does involve the somewhat peculiar technique of artificially raising the
degree of singularity of part of the bare amplitude.  In this case we have a
test of the compatibility of this technique with Ward identities.  The
regulated Ward identities for the analogous form of $D^{\rm sym}_{\mu\nu\rho}$
are discussed in the next section, but we note here that the Ward-like
identity
$$\eqalign{&{\partial\over\partial z_\rho} D_{\mu\nu\rho} (x,y,z) = - \left(
{\partial\over\partial x} + {\partial\over\partial y}\right)_\rho
D_{\mu\nu\rho} (x,y,z) \cr
&= - {1\over 24} \left[\left( \delta_{\mu\nu} \square - \partial_\mu
\partial_\nu\right) \square {\ln M^2(x-y)^2\over (x-y)^2} \right] (x-y)^2
\left(\square_x - \square_y\right) {1\over (x-z)^2(y-z)^2} \cr
&= {\pi^2\over 6} \left( \delta_{\mu\nu} \square - \partial_\mu\partial_\nu
\right) \square {\ln M^2(x-y)^2\over (x-y)^2} \cdot \left( \delta(x-z) -
\delta(y-z)\right) \ \ ,\cr}\eqno(4.10)$$
relates $D_{\mu\nu\rho}$ to a regulated, gauge invariant self-energy.  This
suggests that the divergences of $D_{\mu\nu\rho}$ are indeed controlled by
our procedure.

A more complete proof that the regularization (4.9) is correct requires the
Fourier transform
$$D_{\mu\nu\rho}(p_1,p_2) = \int d^4x\, d^4y\, e^{i(p_1\cdot x + p_2\cdot y)}
D_{\mu\nu\rho}(x,y,0)\ \ .\eqno(4.11)$$
To carry out the $x$ and $y$ integrations we insert
$$(x-y)^2 \left( {\partial\over\partial x_\rho}-{\partial\over\partial
y_\rho}\right) {1\over x^2 y^2} = i \int {d^4k_1\, d^4k_2\over(2\pi)^4} e^{-i
(k_1\cdot x+k_2\cdot y)} \left( {\partial\over \partial k_{1\sigma}} -
{\partial\over \partial k_{2\sigma}}\right)^2 \left[ {\left(
k_1-k_2\right)_\rho
\over k^2_1 \, k^2_2}\right]\eqno(4.12)$$
and
$$\left( \delta_{\mu\nu} \square - \partial_\mu \partial_\nu\right) \square
{\ln M^2(x-y)^2\over (x-y)^2} = -\int {d^4k_3\over (2\pi)^2}
e^{-ik_3\cdot(x-y)} \left( \delta_{\mu\nu}k^2_3 - k_{3\mu} k_{3\nu}\right) \ln
{k^2_3\over \bar{M}^2} $$
where the Fourier transform table of Ref.~[1] has been used.  We then
obtain
$$\eqalign{
D_{\mu\nu\rho}(p_1,p_2) &=  {\pi^2i\over 6} \int d^4k\ln \left( {\left(
k+p_1\right)^2 \over
\bar{M}^2}\right) \cr
&\times\left[ \left( k+p_1\right)^2 \delta_{\mu\nu} - \left(
k+p_1\right)_\mu \left( k+p_1\right)_\nu\right]
\square_k \left[ {\left( 2k+p_1+p_2\right)_\rho \over
k^2 \left( k + p_1 + p_2\right)^2}\right] \ \ .\cr}\eqno(4.13)$$
To test that this somewhat unconventional loop integral is finite, it is
sufficient to examine the leading terms as $k\to\infty$ which are formally of
order $1/k^3$ and $1/k^4$.  These leading terms are
$$\left( \left( k+p_1\right)^2 \delta_{\mu\nu} - \left( k+p_1\right)_\mu
\left( k+p_1\right)_\nu\right) \square_k \left( - {\partial\over \partial
k_\rho} \left( {1\over k^2}\right) - {1\over 2} \left( p_1+p_2\right)_\lambda
{\partial\over\partial k_\lambda}\ {\partial\over\partial k_\rho} \left(
{1\over k^2}\right) \right) \ \ .\eqno(4.14)$$
Since $\square 1/k^2=0$ for $k\not=0$, these vanish identically, so the loop
integral in (4.13) is ultraviolet finite.  Further, shifts in the loop momenta
$k$ are permitted, since the $1/k^3$ term is absent.
If one develops the asymptotic series in $k$ further, one sees that the first
term which contributes to (4.13) has three powers of the external momenta.
Notice that since the
leading term in the integral is proportional to at least the momenta $p_i$,
this is  equivalent in coordinate space to an amplitude which has at least
three
derivatives of a singular function with a well-defined transform.

We now study the conformal tensor $C^{\rm sym}_{\mu\nu\rho}(x,y,z)$.  The
individual terms of (1.9) have an ultraviolet divergent Fourier transform.
The divergence cancels in the cyclic sum of (1.10) but there remains a shift
ambiguity proportional to the bare gluon vertex (1.12).  At first sight it
was surprising to find that $C^{\rm sym}_{\mu\nu\rho}(x,y,z)$ was finite, but
it is actually a direct consequence of the fact that the cutoff-dependent part
of any permutation odd tensor $A^{\rm sym}_{\mu\nu\rho}(x,y,z)$ is proportional
to (1.12) in $p$-space.  Suppose we had picked any pair of permutation odd
conformal tensors, say $A^{\rm sym}_{\mu\nu\rho}(x,y,z)$ and $B^{\rm
sym}_{\mu\nu\rho}(x,y,z)$, rather than $C^{\rm sym}_{\mu\nu\rho}$ and $D^{\rm
sym}_{\mu\nu\rho}$ of (1.10).  Then by examination of the cutoff-dependent
part of $A^{\rm sym}_{\mu\nu\rho}(x,y,z)$ and $B^{\rm
sym}_{\mu\nu\rho}(x,y,z)$, we could select a linear combination with
finite Fourier transform.

The simplest way to confirm that $C^{\rm sym}_{\mu\nu\rho}$ has the properties
stated above is to relate this tensor to the Feynman amplitudes
for the ghost and quark loop
contributions to the three-gluon vertex.  From (3.5) and (3.10), one
obtains
$$f^{abc} C^{\rm sym}_{\mu\nu\rho} (x,y,z) = - \left( 4\pi^2\right)^3 \left\{
{2\over C} \  {\delta^3\hbox{(1.b)} \over \delta B^a_\mu (x) \delta B^b_\nu (y)
\delta B^c_\rho(z)} + {\delta^3\left[ \hbox{(2.a)} + \hbox{(2.b)} \right]\over
\delta B^a_\mu (x) \delta B^b_\nu (y) \delta B^c_\rho(z) }\right\}\ \
.\eqno(4.15)$$
We then use standard $p$-space Feynman rules for these loop graphs.
The result is a loop integral with a formal linear divergence $\int
d^4k\,k_\mu k_\nu k_\rho/k^6$, but no log divergent terms with numerator
quadratic in $k$.

The finiteness of $C^{\rm sym}_{\mu\nu\rho}(x,y,z)$ can also be shown in real
space, without calculation of the Fourier transform.  It is not difficult to
show that each contribution to the cyclic sum (1.10) for $C^{\rm
sym}_{\mu\nu\rho}$ can be expressed in terms of $T_{\mu\nu\rho}$ and
$V_{\mu\nu\rho}$, for example
$$\eqalign{ &C_{\mu\nu\rho}(x,y,z) = - \delta_{\mu\nu} \left(
T_{\lambda\rho\lambda} + T_{\rho\lambda\lambda} - T_{\lambda\lambda\rho}
\right) + {1\over 2} \delta_{\mu\rho} \left( T_{\nu\lambda\lambda} -
T_{\lambda\nu\lambda} + T_{\lambda \lambda\nu}\right) \cr
&- {1\over 2} \delta_{\rho\nu} \left( T_{\mu\lambda\lambda} -
T_{\lambda\mu\lambda} + T_{\lambda\lambda\mu} \right) + 2\left( T_{\mu\rho\nu}
+ T_{\rho\nu\mu}\right) - V_{\nu\rho\mu} ( y-x,  z-x) - V_{\rho\mu\nu}
(z-y,x-y) \ \ .\cr}\eqno(4.16)$$

We then regulate these quantities as in (4.2 -- 4.6), but include different
scale masses $M_c$ for independent traces.  We then find that the overall
scale-dependence cancels in $C^{\rm sym}_{\mu\nu\rho}$ but $\ln M_i/M_j$ terms
multiply the real space form of (1.12) remain.  This is the signal in
differential regularization of a quantity with finite but ambiguous Fourier
transform, and it is very similar to the axial fermion triangle anomaly in
Section~II.D of Ref.~[1].

The regularized form of $C^{\rm sym}_{\mu\nu\rho}(x,y,z)$ found by the
procedure above could be used as the regularized contribution of this tensor
to the three-gluon vertex, but it is a very complicated form and we have found
a much simpler form by combining ideas of conformal invariance, differential
regularization and Ward identities.

One suspects that the tensor $C^{\rm sym}_{\mu\nu\rho}$ satisfies a trivial
bare Ward identity
$${\partial\over\partial z_\rho} C^{\rm sym}_{\mu\nu\rho} (x,y,z) = 0
\eqno(4.17)$$
because the ultraviolet divergence, which would normally be present on both
sides of a non-trivial identity of the form (1.4), has cancelled.  One can
confirm this by using conformal invariance to take the limit$^{10}$ as one
of points goes to $\infty$ (for example, $y_\mu\to\infty$).  It is easy to
compute the $\partial/\partial z_\rho$ derivatives of $C_{\mu\nu\rho}(x,y,z)$
and its cyclic permutations directly from (1.9) in this limit.  The non-local
contribution ($z\not= x)$ vanishes trivially in the cyclic sum, and one also
shows that there is no quasi-local $\delta(z-x)\sum_{\mu\nu} (y)$ term, thus
verifying (4.17).  We call (4.17) a bare Ward identity because the Ward
identity of a regularized form of $C^{\rm sym}_{\mu\mu\rho}$ in general
contains ultra-local terms $\delta(z-x)\left(\partial_\mu\partial_\nu -
\delta_{\mu\nu}\square\right) \delta(x-y)$.  Such terms correspond to the
intrinsic ambiguity of $C^{\rm sym}_{\mu\nu\rho}$, and they cannot be detected
at large $y$.

Motivated by the Ward identity (4.17), we sought a mathematical representation
of a permutation odd tensor with the conformal inversion property (1.5) of a
three-point function of currents in which the trivial gauge property was
carried by explicit derivatives in $x$, $y$, and $z$.  This led us to the {\it
ansatz\/}
$$\eqalignno{
Q_{\mu\nu\rho}(x,y,z) &= \left(\delta_{\mu a} \partial^x_b - \delta_{\mu b}
\partial^x_a \right) \left( \delta_{\nu c} \partial^y_d - \delta_{\nu d}
\partial^y_c \right) \left( \delta_{\rho e} \partial^z_f - \delta_{\rho f}
\partial^z_e \right)\cr
&\times \partial^x_a \partial^y_c \ln (x-y)^2 \partial^x_b \partial^z_e \ln
(x-z)^2 \partial^y_d \partial^z_f \ln (z-y)^2 &(4.18\hbox{a})
\cr\noalign{\vskip 0.2cm}
&= - \left( \delta_{\mu a} \square  - \partial_\mu \partial_a\right)^x \left(
\delta_{\nu b} \square - \partial_\nu \partial_b\right)^y \left( \delta_{\rho
c}\square -\partial_\rho \partial_c\right)^z \cr
&\times \partial^x_a \ln (x-y)^2\partial^y_b \ln (y-z)^2 \partial^z_c\ln
(z-x)^2\ \ . &(4.18\hbox{b}) \cr}$$
To see that (4.18a) has the correct inversion property, we note that for fixed
$y,z$
$$\left( \delta_{\mu a} \partial^x_b - \delta_{\mu b} \partial^x_a \right)
\left[ \partial^x_a \ln (x-y)^2 \partial^x_b \ln (x-z)^2\right] = \partial^x_b
\left[ \partial^x_\mu \ln (x-y)^2 \partial^x_b \ln (x-z)^2 - (b\leftrightarrow
\mu)\right] \ \ .\eqno(4.19)$$
This equation has the structure $j_\mu = \partial_b F_{\mu b}$ where $F_{\mu
b}$ is an anti-symmetric tensor with the same conformal properties as a field
strength.  Thus (4.19) is conformal invariant for the same reason that
Maxwell's equations are conformal invariant, namely $\partial_b F_{\mu b}$ has
the inversion property of a dimension three current.  In (4.18a), this same
property is symmetrically incorporated in all three variables.  The form
(4.18b) is obtained by ``partial integration'' of one derivative in each log
factor, using the gauge property.

The conformal tensor $Q_{\mu\nu\rho}(x,y,z)$ must be a linear combination of
$C^{\rm sym}_{\mu\nu\rho}$ and $D^{\rm sym}_{\mu\nu\rho}$, and the trivial
gauge property suggests that it must be
proportional to $C^{\rm sym}_{\mu\nu\rho}$.
 The derivatives in (4.18) are very tedious to compute, so we employed a
symbolic manipulation program to calculate $Q_{\mu\nu\rho}$ explicitly, and
compare with $C^{\rm sym}_{\mu\nu\rho}$ in (1.9 -- 1.10).  The result is the
equality
$$C^{\rm sym}_{\mu\nu\rho}(x,y,z) = {1\over 16} Q_{\mu\nu\rho}(x,y,z)\ \
.\eqno(4.20)$$
One remaining subtlety arises because the symbolic manipulation program
ignores $\delta(x-y)$ terms which could arise from expressions such as
$\square\left( 1 /(x-y)^2\right)$.  Therefore, we checked by hand calculation
that there are no hidden quasi-local terms, so that the relation (4.20) is
correct.

Because of the many external derivatives in (4.18b), $Q_{\mu\nu\rho}$ can be
regarded as a regularization of $C^{\rm sym}_{\mu\nu\rho}$ because it assigns
a unique Fourier transform, when the partial integration rule of differential
regularization is used.  We find it astonishing that the tensor $C^{\rm
sym}_{\mu\nu\rho}$ whose Fourier transform is linearly divergent by power
counting can be presented in the form (4.18b) whose Fourier transform contains
six powers of external momentum and is therefore represented by
highly convergent loop integrals.

To see that this regularized version of the finite tensor $C^{\rm
sym}_{\mu\nu\rho}$ makes no finite contribution to the Ward
identity we shall compute the Fourier transform of it.
The Fourier transform of $C^{\rm
sym}_{\mu\nu\rho}$ is
$$\eqalign{C^{\rm sym}_{\mu\nu\rho}
&= 2\pi^2 i \left( \delta_{\mu a} p^2_1 - p_{1\mu}
p_{1a} \right) \left( \delta_{\nu b} p^2_2 - p_{2\nu} p_{2b} \right) \left(
\delta_{\rho c} p^2_3 - p_{3\rho} p_{3c}\right)\cr
&\cdot {\partial\over \partial p_{1a}}\ {\partial\over \partial p_{2b}}\left(
{\partial\over \partial p_{1c}} - {\partial\over \partial p_{2c}}\right) \int
{d^4k\over k^2\left( k-p_1\right)^2 \left( k + p_2\right)^2} \cr}\eqno(4.21)$$
where $p_1+p_2 + p_3=0$.  We can see from this representation that as any
momentum component vanishes, $C^{\mu\nu\rho}$ goes to zero so that there is
no contribution to the Ward identity.  It might seem that as $p_2\to 0$ there
would be an infrared singularity of the loop integration at zero but a
careful study of (5.21) for small $p_2$ indicates that it goes smoothly to
zero as $p_2\to 0$.  Therefore
$C^{\rm sym}_{\rm reg}$ gives no contribution to the Ward identity.
\goodbreak
\bigskip
\noindent{\bf V.\quad WARD IDENTITY AND THE MASS SCALE SHIFT}
\medskip
\nobreak
In this section we will consider the Ward identity (2.10) which relates the
three-gluon vertex and the self-energy.  The vertex is a linear combination of
the tensors $D^{\rm sym}_{\mu\nu\rho}(x,y,z)$ and $C^{\rm
sym}_{\mu\nu\rho}(x,y,z)$.  Since
$C^{\rm sym}_{\mu\nu\rho}$ satisfies a trivial Ward identity, our main
task is to study the $\partial/\partial z_\rho$ divergence of $D^{\rm
sym}_{\mu\nu\rho}$.  We will first obtain the bare form of the Ward identity,
and then study the regulated version associated with each of the regulated
forms of $D^{\rm sym}_{\mu\nu\rho}$ discussed in Section~IV.  The purpose is
to show that the proper relation between renormalized vertex $V_{\mu\nu\rho}$
and self-energy $\Sigma_{\mu\nu}$ can be achieved, as in (1.1) -- (1.2) by
specific choice of the mass scale parameter of the differential regularization
procedure.

The bare Ward identity satisfied by $D^{\rm sym}_{\mu\nu\rho}$ is easily
obtained by following the same approach outlined for $C^{\rm
sym}_{\mu\nu\rho}$.  Exploiting conformal invariance, we take advantage of the
algebraic simplifications that occur in the limit as one of the points (we
choose $y_\mu$) goes to $\infty$.  Evaluating derivatives, we find that
$(\partial/\partial z_\rho) D_{\mu\nu\rho}(x,y,z)$ gives only a local
contribution.  Each of the other cyclic permutations in (1.10) contains
non-local terms $(x\not=y$) which cancel in the sum leaving only $\delta(x-z)$.
The net result is
$${\partial\over\partial z_\rho} D^{\rm sym}_{\mu\nu\rho}(x,y,z)
\underarrow{y\to\infty} -12\pi^2 \delta(x-z) \left( \delta_{\mu\nu} - {2y_\mu
y_\nu\over y^2}\right) {1\over y^6}\ \ .\eqno(5.1)$$
{}From this we infer, using $x,\mu\leftrightarrow y,\nu$ permutation symmetry
of
$D^{\rm sym}_{\mu\nu\rho}$ and translation invariance, the bare Ward identity
$$\eqalign{{\partial\over\partial z_\rho}  D^{\rm sym}_{\mu\nu\rho}(x,y,z) &=
\pi^2 \left[ \delta (x-z) - \delta (y-z)\right] \left( \partial_\mu
\partial_\nu - \delta_{\mu\nu} \square \right) {1\over (x-y)^4} \cr
&\equiv \left[ \delta (x-z) - \delta(y-z)\right] \Sigma_{\mu\nu} (x-y) \ \
.\cr}\eqno(5.2)$$

Comparing this result with the Ward identity (1.4) obeyed by current
correlation functions, we see that the coefficients $k$ and $k_1$ in (1.6)
and (1.7) are related by
$$12\pi^2 k_1 = k\eqno(5.3)$$
while $k_2$ in (1.7) is an independent constant, since $C^{\rm
sym}_{\mu\nu\rho}(x,y,z)$ does not contribute to the Ward identity.

The regulated form of the Ward identity is a more delicate matter because it
tests the treatment of the overall singularity at $x\sim y \sim z$ of $D^{\rm
sym}_{\mu\nu\rho}(x,y,z)$.  In any regularization procedure, there is an
unresolved ambiguity which for a linearly divergent quantity with the discrete
symmetries of $D^{\rm sym}_{\mu\nu\rho}(x,y,z)$ is simply a finite multiple of
the bare Yang--Mills vertex. Similarly the ambiguity in a gauge invariant
self-energy function is a multiple of $\left(\partial_\mu \partial_\nu -
\delta_{\mu\nu}\square\right) \delta( x-y)$.  In differential regularization,
these ambiguities are reflected in the dependence of regulated amplitudes on
the mass scales $M$ which are chosen in (4.4).  Thus we study the regulated
form of (5.2) with self-energy scale $M_\Sigma$
$${\partial\over\partial z_\rho} D^{\rm sym}_{\mu\nu\rho}(x,y,z) = {\pi^2\over
4} \left( \delta(x-z) - \delta(y-z) \right) \left( \square \delta_{\mu\nu} -
\partial_\mu \partial_\nu\right) \square {\ln M^2_\Sigma (x-y)^2\over
(x-y)^2}\ \ ,\eqno(5.4)$$
and we require that this be satisfied for both regularizations of $D^{\rm
sym}_{\mu\nu\rho}(x,y,z)$ given in Section~IV.  For the regularized form
(4.7), we use vertex mass scale $M_{V_1}$, and for the form (4.9), the scale
$M_{V_2}$.  In each case we will find that the Ward identity is satisfied, if
relations of the form
$$\ln\left( {M_{V_1}\over M_\Sigma}\right) = a_1\ \ ,\qquad \ln \left(
{M_{V_2}\over M_\Sigma}\right) = a_2 \eqno(5.5)$$
hold.  Since these relations fix the ambiguity in the vertex up to an overall
scale, the two forms (4.7) and (4.9) will then coincide as renormalized
amplitudes if

$$\ln\left( {M_{V_1}\over M_{V_2}}\right) = a_1 - a_2 \eqno(5.6)$$

Since the bare $D^{\rm sym}_{\mu\nu\rho}$ and bare self-energy are properly
related away from coincident points $x=y=z$, it is sufficient to study a
restricted form of (5.4) in order to fix the mass scale ratios of (5.5).  For
the first regulated version (4.7) it is convenient to use the integrated form
of (5.4)
$${\partial\over\partial z_\rho} \int d^4y\,D^{\rm sym}_{\mu\nu\rho}(x,y,z) =
- {\pi^2 \over 4} \left( \square \delta_{\mu\nu} - \partial_\mu
\partial_\nu\right) \square {\ln M^2_\Sigma (x-z)^2
\over (x-z)^2}\ \ .\eqno(5.7)$$
For the regularization (4.9) a Fourier transform of (5.4) is more convenient,
as we discuss below.

Thus our first task is to insert the regulated form (4.7) and its cyclic
permutations in the left-hand side of (5.7), do the integral $d^4y$ and
compute the $\partial/\partial z_\rho$ divergence.  All integrals can be done
using the intermediate results
$$\eqalign{&\int^R d^4y {1\over (x-y)^2}\ {1\over y^2} = - \pi^2 \left[ \ln
{x^2\over R^2} - 1 \right]\cr
&\int^R d^4y {1\over y^4} = - {1\over 4} \int^R d^4y \,\square {\ln M^2
y^2\over y^2} = \pi^2 \left[ \ln M^2 R^2-1\right] \cr
& \left(\partial_\mu \partial_\nu - {1\over 4} \delta_{\mu\nu} \square \right)
\int d^4y {1\over (x-y)^2 y^2} = - \pi^2 \left( \delta_{\mu\nu} - {4x_\mu
x_\nu\over x^2}\right) {1\over x^2} \cr
&\left[ \partial_\mu \partial_\nu \partial_\rho - {1\over 6} \left(
\delta_{\mu\nu} \partial_\rho + \delta_{\mu\rho} \partial_\nu +
\delta_{\nu\rho}\right) \square \right] \int d^4y {1\over (x-y)^2 y^2} = \cr
&= -16\pi^2 \left[ x_\mu x_\nu x_\rho - {1\over 6} x^2 \left( \delta_{\mu\nu}
x_\rho + \delta_{\nu\rho} x_\mu + \delta_{\rho\mu} x_\nu\right) \right]
{1\over x^6} \cr}\eqno(5.8)$$
An infrared cutoff $R$ at some large value of $y$ is required because
individual terms in the contribution of the permutation
$D_{\rho\mu\nu}(z,x,y)$ to (5.7) are infrared divergent, although this
divergence cancels in the full contribution.  The last two integrals are
obtained by explicit differentiation of the first result.  The second integral
is evaluated using the differential  regularization recipe in which the
singular contribution for small $y$ is ignored.  The large $y$ contribution is
obtained from the divergence theorem.  One uses translational invariance to
replace $x\to x-z$ in applying (5.8) to (5.7).

We record the results of the three cyclic permutations:
$$\eqalign{
{\partial\over \partial z_\rho} \int d^4y D_{\mu\nu\rho} (x,y,z) &= {\pi^2
\over 6} \left( \partial_\mu \partial_\nu - \delta_{\mu\nu} \square \right)
\square {\ln M^2_{V_1} (x-y)^2\over (x-z)^2} \cr
&+ {\pi^4\over 9} \left(
\partial_\mu \partial_\nu - {1\over 4} \delta_{\mu\nu} \square \right)
\delta(x-z) \cr
{\partial\over \partial z_\rho} \int d^4y\, D_{\nu\rho\mu} (y,z,x) &= -
{\pi^2\over 12} \left( \partial_\mu \partial_\nu - \delta_{\mu\nu}\square
\right) \square {\ln M^2_{V_1} (x-y)^2\over (x-z)^2} \cr
&+ {\pi^4\over 9} \left( -
{5\over 4} \partial_\mu \partial_\nu + 2 \delta_{\mu\nu} \square \right)
\delta(x-z) \cr
{\partial\over\partial z_\rho} \int d^4y\, D_{\rho\mu\nu}(z,x,y) &=
{\pi^2\over 6} \left( \partial_\mu \partial_\nu - \delta_{\mu\nu} \square
\right) \square {\ln M^2_{V_1} (x-z)^2\over (x-z)^2}
\cr
&- {\pi^4\over 9} \left(
{1\over 2} \partial_\mu \partial_\nu + \delta_{\mu\nu} \square \right)
\delta(x-z) \ \ .\cr}\eqno(5.9)$$
We add these contributions to obtain the final gauge invariant result
$${\partial\over\partial z_\rho} \int d^4y\, D^{\rm sym}_{\mu\nu\rho}(x,y,z) =
- {\pi^2\over 4} \left( \square \delta_{\mu\nu} - \partial_\mu
\partial_\nu\right) \square \left[ {\ln M^2_{V_1} (x-z)^2\over (x-z)^2} +
{1\over 12} \  {1\over (x-z)^2}\right]\ \ ,\eqno(5.10)$$
which, when compared with (5.7) yields the mass scale relation
$$\ln {M^2_{V_1}\over M^2_\Sigma} = - {1\over 12} \ \ .\eqno(5.11)$$

We next consider the second regularized form (4.9) of $D^{\rm
sym}_{\mu\nu\rho}$.  Since the Fourier transform (4.13) is quite simple, we
choose to work in momentum space.  The Fourier transform of (5.4) is
$$i \left( p_1+p_2\right)_\rho D^{\rm sym}_{\mu\nu\rho} \left( p_1,p_2\right)
= \Sigma_{\mu\nu} \left( p_2\right) - \Sigma_{\mu\nu} \left( p_1\right)\ \
.\eqno(5.12)$$
The restriction we use to fix mass scales is the analogue of the original Ward
identity of quantum electrodynamics and is obtained by applying
$\partial/\partial p_{2\lambda}$ to (5.12), and then setting $p_2 = - p_1= -
p$,
which is equivalent to $p_3=0$.  The result is
$$\eqalign{
i D^{\rm sym}_{\mu\nu\lambda} (p,-p) &= - {\partial\over\partial p_\lambda}
\Sigma_{\mu\nu} (p) \cr
&= \pi^4 {\partial\over\partial p_\lambda} \left[ \left( p^2 \delta_{\mu\nu} -
p_\mu p_\nu\right) \ln {p^2\over \bar{M}^2_\Sigma}\right] \cr}\eqno(5.13)$$
where we have used (5.4) and the Fourier transform result of Appendix A of
Ref.~[1] to obtain the last line.  Note that $\bar{M}_\Sigma =
2M_\Sigma/\gamma$ where $\gamma=1.781$ is Euler's constant.

{}From (4.11) and permutations, we find that
$$D^{\rm sym}_{\mu\nu\lambda} (p,-p) = D_{\mu\nu\lambda} (p,-p) +
D_{\nu\lambda\mu} (-p,0) + D_{\lambda\mu\nu} (0,p) \ \ .\eqno(5.14)$$

To evaluate $D_{\mu\nu\lambda}(p,-p)$ is simple; we have already shown that
this term obeys a kind of Ward identity by itself.  One finds directly from
(4.13) using
$$\square_p {\partial\over\partial p_\lambda} \left( {1\over p^2}\right) = -
4\pi^2 {\partial\over\partial p_\lambda}\delta(p)\ \ ,$$
$$D_{\mu\nu\lambda}(p,-p) = - {2\pi^4i\over 3}\  {\partial\over\partial
p_\lambda} \left[\left( p^2 \delta_{\mu\nu} - p_\mu p_\nu\right) \ln {p^2\over
\bar{M}^2_{V_2}}\right]\ \ .\eqno(5.15)$$
Next we compute the remaining terms in (5.14) where an elementary integral
must be evaluated, namely
$$\eqalign{\int &d^4k\, \ln \left( {k^2\over \bar{M}^2_{V_2}} \right)\left( k^2
\delta_{\mu\nu} - k_\mu k_\nu \right) \square_k \left( {\left(
2k-p\right)_\lambda\over k^2(k-p)^2}\right) \cr
&= - 2\pi^2 {p_\lambda p_\mu
p_\nu\over p^2} + p_\lambda \delta_{\mu\nu} \left( - 4\pi^2 \ln {p^2\over
\bar{M}^2_{V_2}}\right)
- \left( p_\mu \delta_{\lambda\nu} + p_\nu \delta_{\lambda\mu} \right) \left(
2\pi^2 \ln {p^2\over \bar{M}^2_{V_2}} + \pi^2\right) \cr}\eqno(5.16)$$
Thus
$$D_{\nu\lambda\mu} (-p,0) + D_{\lambda\mu\nu} (0,p) = \left(-{\pi^4i\over
3}\right) {\partial\over \partial p_\lambda} \left( p^2 \delta_{\mu\nu}-p_\mu
p_\nu\right) \left( \ln {p^2 \over \bar{M}^2_{V_2}} - {1\over 2}\right)
\eqno(5.17)$$
and so we obtain
$$D^{\rm sym}_{\mu\nu\lambda} (p,-p) = \left( - \pi^4i\right) {\partial\over
\partial p_\lambda} \left( p^2 \delta_{\mu\nu} - p_\mu p_\nu\right) \left( \ln
{p^2 \over \bar{M}^2_{V_2}} - {1\over 6}\right) \eqno(5.18)$$

We see that the only difference between (5.18) and (5.13) is a linear term in
$p$ which is the expected local violation of the Ward identity due to
regularization ambiguities.  The Ward identity is satisfied exactly if we fix
the mass scale ratio
$$\ln \left({M^2_{V_2}\over M^2_\Sigma}\right) = - {1\over 6} \ \
.\eqno(5.19)$$
Using (5.11) we see that
$$\ln \left({M^2_{V_1}\over M^2_{V_2}}\right) = - {1\over 12} + {1\over 6} =
{1\over 12}\ \ .\eqno(5.20)$$
This calculation illustrates the ``robustness'' of the differential
regularization method.  Two rather different regularizations of the same
amplitude are simply related by a proper choice of scale parameters.
\goodbreak
\bigskip
\noindent{\bf VI.\quad THE BETA-FUNCTION}
\medskip
Although the one-loop $\beta(g)$ for non-Abelian gauge theories was calculated
in Ref.~[1] using differential regularization, we shall recompute it here from
the present viewpoint which emphasizes our principal result that the gluon
vertex function is the linear combination of conformal tensors given in
(3.12).

We use the notation (7.1) and the renormalization group equation (7.4), which
requires that the classical and regulated one-loop contributions to the vertex
function, denoted respectively by $\Gamma^0_{\mu\nu\rho}$ and
$\Gamma^1_{\mu\nu\rho}$ are related by
$$\eqalign{M&{\partial\over\partial M}\Gamma^1_{\mu\nu\rho} (x,y,z) = -
\beta(g) {\partial\over\partial g}\Gamma^0_{\mu\nu\rho} (x,y,z) \cr
&= 2 {\beta(g)\over g^3} \left\{ \delta_{\mu\nu} \left( \partial^x_\rho -
\partial^y_\rho\right) + \delta_{\nu\rho}\left( \partial^y_\mu -
\partial^z_\mu\right) + \delta_{\rho\mu}\left( \partial^z_\nu -
\partial^x_\nu\right)\right\} \delta(x-y) \delta(y-z) \cr}\eqno(6.1)$$
where the classical term is easily computed directly from $S[B]$ in (2.1).
Using (3.12) and the fact that only the regulated tensor $D^{\rm
sym}_{\mu\nu\rho}$ is scale-dependent, we see that the left-hand side of (6.1)
is simply
$$M{\partial\over\partial M} \Gamma^1_{\mu\nu\rho} (x,y,z) =  {1\over 48\pi^6}
\left( 11C - 2N_f\right) M{\partial\over\partial M} D^{\rm sym}_{\mu\nu\rho}
(x,y,z)\ \ .\eqno(6.2)$$

It is rather trivial to compute the scale derivative of the regularized form
(4.7) of $D_{\mu\nu\rho}$ and add permutations to obtain
$$M{\partial\over \partial M} D^{\rm sym}_{\mu\nu\rho} (x,y,z) = -2\pi^4
\left\{ \delta_{\mu\nu} \left( \partial^x_\rho - \partial^y_\rho\right) +
\delta_{\nu\rho} \left( \partial^y_\mu - \partial^z_\mu\right) +
\delta_{\rho\mu} \left( \partial^z_\nu - \partial^x_\nu\right) \right\}
\delta(x-y) \delta(y-z).\eqno(6.3)$$
{}From (6.1) -- (6.3) one immediately finds the well-known result
$$\beta(g) = - {g^3\over 48\pi^2} \left( 11C - 2N_f\right) + {\cal O} (g^5) \
\ .\eqno(6.4)$$

It is a useful test of the ideas underlying the second regularization of
$D_{\mu\nu\rho}$ in (4.9) to see that the same result can be obtained from
this form.  The scale derivative of (4.9) gives an expression which is
difficult to interpret as a product of delta functions unless an integral with
a smooth function is performed, so it is natural to study the momentum form
(4.13), where the scale derivative is
$$M{\partial\over \partial M} D_{\mu\nu\rho} (p_1,p_2) = -{i\pi^2\over 3} \int
d^4k \left( k^2 \delta_{\mu\nu} - k_\mu k_\nu\right) \square_k {\left( 2k -
p_1 + p_2\right)_\rho\over \left( k - p_1\right)^2 \left( k + p_2\right)^2}
\eqno(6.5)$$
where we have made the (permitted) shift $k\to k-p_1$ of the loop momentum in
(4.13).  Dimensional and symmetry arguments tell us that the integral must
have the form
$$M{\partial\over\partial M} D_{\mu\nu\rho} (p_1,p_2) = A\left(
p_1-p_2\right)_\rho \delta_{\mu\nu} + B \left[ \left( p_1-p_2\right)_\mu
\delta_{\nu\rho} + \left( p_1-p_2\right)_\nu \delta_{\mu\rho}\right]
\eqno(6.6)$$
where $A$ and $B$ are purely numerical constants.  To compute $A$ and $B$ it
is sufficient to evaluate (6.5) with $p_2=-p_1$.  This leads to
$$Ap_\rho \delta_{\mu\nu} + B\left[ p_\mu \delta_{\nu\rho} + p_\nu
\delta_{\mu\rho}\right] = -{i\pi^2\over 3} \int d^4k \left( k_2 \delta_{\mu\nu}
- k_\mu k_\nu\right) \square_k {(k-p)_\rho\over (k-p)^4} \ \ .\eqno(6.7)$$
In this form the integral is elementary, since
$$\square_k {(k-p)_\rho\over (k-p)^4} = {1\over 2} \  {\partial\over \partial
p_\rho}\square_k {1\over (k-p)^2} = - 2\pi^2 {\partial\over \partial p_\rho}
\delta(k-p) \eqno(6.8)$$
so
$$Ap_\rho \delta_{\mu\nu} + B\left[ p_\mu \delta_{\nu\rho} + p_\nu
\delta_{\mu\rho}\right] = {i2\pi^4\over 3} \  {\partial\over\partial p_\rho}
\left( p^2 \delta_{\mu\nu} - p_\mu p_\nu\right) \eqno(6.9)$$
which gives
$$A =  {i4\pi^4\over 3}\ \ ,\qquad B =- {i2\pi^4\over 3}\ \ .\eqno(6.10)$$

We insert this into (6.6) and add cyclic permutations to obtain
$$\eqalign{
M{\partial\over \partial M} D^{\rm sym}_{\mu\nu\rho} \left( p_1,p_2,p_3
\right) &= M{\partial\over\partial M} \left[ D_{\mu\nu\rho} \left( p_1,
p_2\right) \right.\cr
&\left. + D_{\nu\rho\mu} \left( p_2, p_3 = - (p_1+p_2)\right) +
D_{\rho\mu\nu} \left( p_3=-(p_1+p_2),p_1\right)\right] \cr
&=  i 2\pi^4 \left\{ \delta_{\mu\nu} \left( p_1-p_2\right)_\rho +
\delta_{\nu\rho} \left( p_2-p_3\right)_\mu + \delta_{\rho\mu} \left(
p_1-p_2\right)_\nu\right\} \ \ .\cr}\eqno(6.11)$$
If this result is inserted in the Fourier transform of (6.2) and (6.1), we
again find the beta-function (6.4).
\goodbreak
\bigskip
\noindent{\bf \bf VII.\quad OUTLOOK BEYOND ONE-LOOP}
\medskip
\nobreak
Our approach to the three-gluon vertex has been largely ``experimental,'' and
we do not yet have a theoretical explanation of the gauge-specific conformal
property found at one-loop order.  Nevertheless it is of some interest to
consider the possible role of conformal symmetry beyond one-loop.  We discuss
here an admittedly speculative scenario based on the interplay of
renormalization group equations and Ward identities.  We will see that exact
conformal invariance cannot hold in higher order because of the twin problems
of subdivergences and gauge-dependence.  However, one might have a situation
in which, in a given order of perturbation theory, the three-gluon vertex or
the three-point current correlation function has a certain conformal invariant
primitive core which is a linear combination of $D^{\rm sym}_{\mu\nu\rho}$ and
$C^{\rm sym}_{\mu\nu\rho}$ plus conformal breaking terms which are determined
by the renormalization group in terms of lower-order amplitudes.

To simplify notation let us define 1PI two- and three-point functions in the
background field formalism as
$$\eqalign{{\delta^2\over \delta B^a_\mu (x) \delta B^b_\nu(y)} \left( S[B] +
\Omega [B,J]_{j=0}\right) &\equiv \delta^{ab} \Gamma_{\mu\nu} (x-y) \cr
{\delta^3\over \delta B^a_\mu(x) \delta B^b_\nu(y) \delta B^c_\rho(z)} \left(
S[B] + \Omega [B,J]_{j=0} \right) &\equiv f^{abc} \Gamma_{\mu\nu\rho} (x,y,z)
\ \ .\cr}\eqno(7.1)$$
The Ward identity (2.10) then becomes
$${\partial\over \partial z_\rho} \Gamma_{\mu\nu\rho} (x,y,z) = \left[
\delta(y-z) - \delta(x-z) \right] \Gamma_{\mu\nu} (x-y) \eqno(7.2)$$

We use the results of Kluberg--Stern and Zuber$^{11}$ and
Abbott$^{6}$ on the renormalization properties of the background method
to determine the renormalization group equations satisfied by
$\Gamma_{\mu\nu}$ and $\Gamma_{\mu\nu\rho}$.  First we note that our
background field $B^a_\mu(x)$ has no anomalous dimension to all orders in
perturbation theory because $B^a_\mu(x)$ is related to $A^a_\mu(x)$ of
Refs.~[6] and [11] by
$$\eqalign{ B^a_\mu(x) &= g_{\rm bare} \, A^a_\mu(x)_{\rm bare} \cr
&= g\,Z_g\sqrt{Z_A}\, A^a_\mu (x)_{\rm ren} \cr
&= g\,A^a_\mu (x)_{\rm ren}\cr}\eqno(7.3)$$
since $Z_g\sqrt{Z_A} = 1$.  It is also known that
renormalization of the gauge parameter $a$ is required to make the two-point
function of the quantum field $b^a_\mu$ multiplicatively renormalizable in any
gauge except Landau gauge.  These facts suggest that the renormalization group
equations take the form
$$\eqalignno{\left[ M {\partial\over\partial M} + \beta(g) {\partial\over
\partial g} + \delta(g) a{\partial\over\partial a}\right] \Gamma_{\mu\nu}
(x-y) &= 0 &(7.4) \cr
\left[ M {\partial\over \partial M} + \beta(g) + \delta(g) a {\partial\over
\partial a}\right] \Gamma_{\mu\nu\rho} (x,y,z) &= 0 &(7.5)\cr}$$
where $\beta(g) = g^3 \beta_1 + g^5 \beta_2 + \ldots$ and $\delta(g) = g^2
\delta_1 + g^4 \delta_2 + \ldots$ and the subscripts 1 and 2 are the
loop-order of the expansion coefficients of $\beta(g)$ and $\delta(g)$.  We
remind readers that in our conventions the $\ell$-loop contributions to
$\Gamma_{\mu\nu}$ and $\Gamma_{\mu\nu\rho}$ carry the power $g^{2\ell-2}$.

Although (7.4) -- (7.5) are generally valid we use them here only for
separated points.  Therefore the classical contributions can be dropped, and
the differential regularization of the overall singularity in a given order of
perturbation theory is irrelevant.  Thus the result of Section~II.G of
Ref.~[1] for $\Gamma_{\mu\nu}$ in one-loop order can be written as
$$\Gamma_{\mu\nu} (x) =- {\beta(g)\over \pi^2 g^3} \left( \delta_{\mu\nu}
\square - \partial_\mu \partial_\nu\right) {1\over x^4} \ \ ,\qquad x\not=0\ \
.\eqno(7.6)$$
Let us bring in some information about higher-order terms in
$\Gamma_{\mu\nu}$.  Using the structure of (7.4) and the fact that $\beta_1$
and $\beta_2$ are gauge-independent, one sees that $\Gamma_{\mu\nu}$ is
described by
$$\Gamma_{\mu\nu}(x) = - {1\over \pi^2} \left( \delta_{\mu\nu} \square -
\partial_\mu\partial_\nu\right) \left[ {1\over x^4} \left( {\beta(g)\over g^3}
- g^4 \beta_1\beta_2 \ln M^2 x^2\right)\right] \eqno(7.7)$$
through three-loop order.  The last term is the result of uncancelled
subdivergences in three-loop order, and the only allowed gauge dependence in
(7.7) is the coefficient $\beta_3$.

What can be said about $\Gamma_{\mu\nu\rho}$?  First let us refer to the
calculation of the linear deviation from the Feynman gauge in Appendix A, and
denote by $R_{\mu\nu\rho}(x,y,z)$ the variational derivative of the final
result (A.7) which is a one-loop contribution to ${\partial\over\partial a}
\Gamma^1_{\mu\nu\rho}$ at $a=1$.  Note also that $R_{\mu\nu\rho}$ satisfies a
trivial Ward identity, {\it e.g.\/} $\partial^z_\rho R_{\mu\nu\rho}(x,y,z)=0$.
 It then follows from (7.5) that the two-loop contribution to the vertex
function $\Gamma^2_{\mu\nu\rho}$ in Feynman gauge satisfies
$$M{\partial\over\partial M} \Gamma^2_{\mu\nu\rho} = - g^2\delta_1
R_{\mu\nu\rho}\ \ .\eqno(7.8)$$
We shall write down the following solution of (7.8)
$$\Gamma^2_{\mu\nu\rho} (x,y,z) = {1\over 6} g^2\left\{
 \delta_1 \ln \left[ - \left(
\square_x\square_y\square_x\right)\big/ M^6\right] R_{\mu\nu\rho} (x,y,z) +
N_{\mu\nu\rho} (x,y,z)\right\} \eqno(7.9)$$
where $N_{\mu\nu\rho}$ is a permutation odd tensor which is independent of
scale $M$.  This solution may not be unique, but it does illustrate one way in
which our conformal scenario can work.  Note that a small change in the scale
parameter corresponds to a perturbative correction to the Feynman gauge
condition, $a=1 + {\cal O}(g^2)$, and especially that the scale-dependent term
satisfies a trivial Ward identity.  Thus the only {\it a priori\/}  constraint
on $N_{\mu\nu\rho}$ is that it satisfies the Ward identity (7.2) with the
two-loop contribution $\Gamma^2_{\mu\nu}$ on the right-hand side.  One
solution of this is the conformal tensor
$$N_{\mu\nu\rho} = -{1\over \pi^4}\beta_2 D^{\rm sym}_{\mu\nu\rho} + \gamma_2
C^{\rm sym}_{\mu\nu\rho} \eqno(7.10)$$
where we have used (5.3), and the constant $\gamma_2$ is undetermined because
$C^{\rm sym}_{\mu\nu\rho}$ satisfies a trivial Ward identity. The analysis
presented here cannot substitute for the very difficult job of a complete
two-loop calculation, yet it incorporates all the general properties
which the true amplitudes must satisfy.  The tensor $N_{\mu\nu\rho}$ could be
the conformal invariant core of the two-loop vertex function.

Note that it is the fact that $\Gamma^1_{\mu\nu\rho}$ is gauge-dependent that
forces the scale-dependence in (7.9) and indicates that subdivergences do not
cancel in $\Gamma^2_{\mu\nu\rho}$.  The situation would be the same even if the
linear deviation from Feynman gauge was conformal invariant, so that
$R_{\mu\nu\rho}\sim C^{\rm sym}_{\mu\nu\rho}$.  In the Landau gauge, $a=0$,
the problematic $\partial/\partial a$ term in (7.5) disappears, and there can
be no subdivergences in $\Gamma^2_{\mu\nu\rho}$.  However the calculations
described in Appendix~C indicate that $\Gamma^1_{\mu\nu\rho}$ is not conformal
invariant in Landau gauge.  It is still possible that $\Gamma^2$ takes the
conformal form (7.10) in this gauge, but this does not seem to be interesting.

As a separate question, one can also study the conformal properties of gauge
invariant operators such as the color singlet currents $J_{\mu AB} = \bar{q}_A
\gamma_\mu q_B$ in a colored quark theory where the quark fields $q_A(x)$ are
labelled by explicit flavor indices $A$ (with color and spin labels
suppressed).  The three-point function of such currents should have similar
properties to that of the $SU(2)$ currents $J^a_\mu = {1\over 2} \bar{\psi}_i
\tau^a_{ij} \gamma_\mu \psi_j$ in the model discussed by Baker and Johnson
in which each field $\psi_i(x)$, with $i=1,2$, is coupled to an Abelian gauge
field.  In both cases, current correlation functions are independent of the
gauge condition chosen for the internal gauge field.  Since, as we saw in
Section~III, a conformal transformation can be compensated by a gauge
transformation, the correlators of the $J^a_\mu$ have the property that
conformal invariance may be broken by the renormalization procedure, but
gauge-fixing is not a problem.

Let us factor out fermion flavor indices as in (7.1) and use a notation
in which the space-time part of the two- and three-point current correlators
are denoted by $\hat\Gamma_{\mu\nu}(x-y)$ and
$\hat\Gamma_{\mu\nu\rho}(x,y,z)$.  For non-coincident points, these
gauge-independent amplitudes obey renormalization group equations of the
simple form
$$\eqalignno{
\left[ M {\partial\over\partial M} + \beta(g) {\partial\over \partial
g}\right] \hat\Gamma_{\mu\nu} (x-y) &= 0 &(7.11) \cr
\left[ M{\partial\over\partial M} + \beta(g) {\partial\over\partial g}\right]
\hat\Gamma_{\mu\nu\rho} (x,y,z) &= 0 &(7.12)\cr}$$
and the Ward identity (7.2) holds.  One can see that $\beta(g) {\partial\over
\partial g}$ acts on $\hat\Gamma^2_{\mu\nu}$ as $\hat \Gamma^2_{\mu\nu\rho}$
to produce terms of order $g^4$ in these equations, thus showing that
$\hat\Gamma^3_{\mu\nu}$ and $\hat\Gamma^3_{\mu\nu\rho}$ are scale-dependent
because of subdivergences.  However these same equations show that
subdivergences cancel in two-loop contributions.

To proceed further, we discuss the Abelian model,$^{10}$ although we expect
that the current correlators of the colored quark theory are similar.  From
the work of DeRafael and Rosner,$^{15}$ one can see that
$\hat\Gamma_{\mu\nu}(x)$ has the form (7.7) through three--loop order, where
$\beta(g)$ is the beta-function of quantum electrodynamics.  The results of
Baker and Johnson show that the vertex $\hat\Gamma_{\mu\nu\rho}$ is conformal
invariant through two-loop order and can be expressed as
$$\hat\Gamma_{\mu\nu\rho} (x,y,z) = -{1\over\pi^4}
 \left[ \left( b_1 + b_2 g^2\right)
D^{\rm sym}_{\mu\nu\rho}(x,y,z) + \left( a_1 + a_2 g^2\right) C^{\rm
sym}_{\mu\nu\rho}(x,y,z)\right] + {\cal O}(g^4) \eqno(7.13)$$
where $a_1$ and $a_2$ are numbers whose exact values are not relevant here.

We represent the unknown three-loop contribution as the sum of a
scale-dependent part and a primitive core
$$\hat\Gamma^3_{\mu\nu\rho} (x,y,z) = g^4 \left[ S^3_{\mu\nu\rho} (x,y,z,M) +
N^3_{\mu\nu\rho} (x,y,z)\right] \ \ .\eqno(7.14)$$
Then $S^3_{\mu\nu\rho}$ must satisfy the twin constraints
$$\eqalignno{
M{\partial\over \partial M} S^3_{\mu\nu\rho} &={2\over \pi^4}
 b_1 \left[ b_2 D^{\rm
sym}_{\mu\nu\rho} + a_2 C^{\rm sym}_{\mu\nu\rho}\right] &(7.15) \cr
{\partial\over\partial z_\rho} S^3_{\mu\nu\rho} (x,y,z,M) &= - {b_1b_2\over
\pi^2} \left[ \delta(z-x) - \delta(z-y)\right] \left( \delta_{\mu\nu}\square
- \partial_\mu \partial_\nu\right) {\ln M^2 (x-y)^2\over (x-y)^4}\ \ . &(7.16)
\cr}$$
It is plausible that a combined solution of (7.15) -- (7.16) can be obtained.
Then the tensor $N^3_{\mu\nu\rho}$ is constrained only by the simple Ward
identity for which one solution is the conformal tensor
$$N^3_{\mu\nu\rho} = -{1\over \pi^4} b_3 D^{\rm sym}_{\mu\nu\rho} + \gamma_3
C^{\rm sym}_{\mu\nu\rho}\ \ ,\eqno(7.16)$$
which would represent the primitive conformal core of the three-loop vertex
function.

The last topic to be discussed here is an idea for an unconventional gauge
fixing and renormalization procedure for gauge theories, which appears to lead
to renormalization group equations without the troublesome
$a{\partial\over\partial a}$ term of (7.4) -- (7.5).  Further study is
certainly
required to see if this procedure is consistent.  If so, then exact conformal
invariance can hold at two-loop order for the background gluon vertex
function, and the situation in higher-order becomes similar to that of the
current correlation function discussed above.

The need to renormalize the gauge parameter $a$ appears first in the one-loop
quantum two-point function $\Gamma^b_{\mu\nu}(x)$.  In momentum space the
conventional renormalized amplitude is$^{16}$
$$\Gamma^b_{\mu\nu} (p) = p^2 \delta_{\mu\nu} + \left( {1\over a}-1\right)
p_\mu p_\nu + g^2 \left[ c_1 + c_2 (a-1)\right] \left( p^2\delta_{\mu\nu} -
p_\mu p_\nu\right) \ln {p^2\over \bar{M}^2} \eqno(7.17)$$
where $c_1$ and $c_2$ are known numerical constants.  This amplitude was
obtained via dimensional regularization, but differential regularization would
give the same result without the appearance of explicit divergences and
counter terms.  Because the one-loop (order $g^2$) contribution is transverse,
this amplitude satisfies the renormalization group equation (to one-loop
order)
$$\left[ M {\partial\over \partial M} + \delta(g) a{\partial\over\partial a} -
2\gamma (g) \right] \Gamma^b_{\mu\nu} (p) = 0 \eqno(7.18)$$
with
$$\gamma(g) = - {1\over 2} \delta(g) = - g^2 \left[ c_1 + c_2 (a-1) \right]
\eqno(7.19)$$
and this is the result expected from the renormalized Lagrangian of
Kluberg--Stern and Zuber (see Appendix ~B).

We now consider the combination of a conventional and a new non-local gauge
fixing term
$$\widetilde S_{gf}
= {1\over 2a} \int d^4x(D\cdot b)^2 + {\alpha g^2\over 8\pi^2}
\int d^4x\,d^4y\,D\cdot b(x) \square {\ln M^2(x-y)^2\over (x-y)^2} D\cdot b(y)
\ \ .\eqno(7.20)$$
The last integral is finite at $x\approx y$ if interpreted with the partial
integration rule of differential regularization.  The last term gives an
additional contribution to the momentum space two-point function proportional
to $p_\mu p_\nu \ln p^2/\bar{M}^2$ as we can see from the Fourier transform
rule of Appendix~A of Ref.~[1].  The new renormalized two-point function is
$$\widetilde\Gamma^b_{\mu\nu}(p)
= \Gamma^b_{\mu\nu}(p) + \alpha g^2 p_\mu p_\nu \ln
{p^2\over \bar{M}^2}\ \ .\eqno(7.21)$$
We now choose $\alpha = [c_1+c_2(a-1)]/a$ and obtain
$$\widetilde\Gamma^b_{\mu\nu}
 (p) = \left\{ 1 + g^2 \left[ c_1 + c_2 (a-1) \right]
\ln {p^2\over \bar{M}^2} \right\} \left[ p^2 \delta_{\mu\nu} + \left( {1\over
a} - 1 \right) p_\mu p_\nu \right] \ \ .\eqno(7.22)$$
This amplitude satisfies the simpler renormalization group equation (to order
$g^2$):
$$\left[ M {\partial\over\partial M} - 2\gamma(g) \right]
\widetilde\Gamma^b_{\mu\nu}
(p) = 0 \eqno(7.23)$$
with $\gamma(g)$ as given in (7.19).

As in Ref.~[1], this equation is obtained essentially by inspection.  The
standard method of deriving renormalization group equations from the cutoff
dependence of renormalization constants can also be implemented in
differential regularization$^{17}$
by introducing a short-distance cutoff and showing
that the surface terms usually neglected in the Fourier transform are actually
cancelled by counterterms.  In the present situation there is a transverse
counterterm $\delta Z{1\over 2} \int d^4x\, b_\mu \left[ \square
\delta_{\mu\nu} - \partial_\mu \partial_\nu\right] b_\nu$ associated with the
one-loop contribution to (7.17), and it seems clear that the surface term
associated with the non-local part of (7.20) has the local form $\delta
Z'\int d^4x{1\over 2} (\partial\cdot b)^2$ with $\delta Z = \delta Z'$ if the
relation $\alpha \left[ c_1+c_2 (a-1)\right]/a$ is enforced.  Thus one would
find a net wavefunction renormalization of the local kinetic terms of $b_\mu$
in agreement with (7.23).

We now discuss the application of (7.20) in the background formalism.  As
written, however, it is not applicable because the non-local term is not
background gauge invariant.  But it can be covariantized if we make the
replacement
$${1\over 4\pi^2} \square {\ln (x-y)^2 M^2\over (x-y)^2}\to \ln \left( - D_\mu
D_\mu/\bar{M}^2\right) \eqno(7.24)$$
where $D_\mu$ is the background covariant derivative in the adjoint
representation.  When $B_\mu=0$, the left- and right-hand sides of (7.24)
coincide, as can be seen by Fourier transformation.  It appears that the usual
argument$^{7,\,8}$ that the background field method gives the correct
$S$-matrix can be extended to cover a non-local $B_\mu$-dependent gauge fixing
term, and we proceed to analyze the implications for the three-gluon vertex.

Let us denote by $\widetilde\Gamma_{\mu\nu\rho}$ the background field vertex
function calculated with the new gauge fixing action (7.20).  We will discuss
only the two-loop contribution
$\widetilde\Gamma^2_{\mu\nu\rho}$ which is of order
$g^2$, and the lowest order in which the non-local part of (7.20) has an
effect. We assume that the appropriate renormalization group equation is the
same as (7.12).  The question of an anomalous dimension for $B_\mu(x)$ in
the new procedure may need reexamination, but it is certain that the one-loop
anomalous dimension vanishes which is sufficient to allow us to examine the
consequences of (7.20) at two-loop order away from the coincident point
singularity.  The $\beta(g) {\partial\over\partial g}$ term makes no
contribution to to order $g^2$, and we find only the simple condition
$$M{\partial\over\partial M}\widetilde\Gamma^2_{\mu\nu\rho} (x,y,z) = 0
\eqno(7.25)$$
which has the direct interpretation that subdivergences cancel among all the
graphical contributions to $\widetilde\Gamma^2_{\mu\nu\rho}(x,y,z)$.  A
complete
two-loop calculation of
$\widetilde\Gamma^2_{\mu\nu\rho}$ is very difficult, but the
cancellation of subdivergences is a computationally simpler question,
 and it is a useful test of the
idea under discussion.  A positive result in no way guarantees that
$\widetilde\Gamma^2_{\mu\nu\rho}(x,y,z)$ is conformal
invariant in the Feynman gauge
$a=1$ or a perturbative modification of this gauge, but there is no reason why
this cannot be the case.

The non-local gauge fixing term discussed above removed the
$a{\partial\over\partial a}$ term from the renormalization group equation only
to one-loop order for $\widetilde\Gamma_{\mu\nu}$ and to two-loop order for
$\widetilde\Gamma_{\mu\nu\rho}$.
It seems plausible that the procedure works quite
generally if the non-local term in (7.20) is chosen appropriately as a
series expansion in $g^2$, and this is an interesting subject for further
investigation.

The conformal scenario we have outlined in this section is very far from
proven, but it does indicate a way in which the conformal property can combine
with the renormalization group and Ward identities to determine the full
structure of vertex functions.
\goodbreak
\bigskip
\noindent{\bf VIII.\quad CONCLUSIONS}
\medskip
\nobreak
In this section we present a partial review of the results of this work, in
which the logical relation of the ideas underlying the conformal property is
emphasized. Conformal invariance holds for renormalized amplitudes only when
the twin difficulties of renormalization scale effects and gauge fixing can be
circumvented. The first requires that we study real space amplitudes away from
coincident points and choose amplitudes in which subdivergences cancel.  The
second problem is avoided for gauge-independent correlation functions.
However, the standard renormalization program in gauge theories requires
consideration of gauge-dependent amplitudes.  Here the situation is somewhat
different for two-, three- and four-point correlation functions.

In gauge field theory, self-energies and two-point current correlators are
constrained by gauge invariance and canonical dimension to have the structure
$$\left( \square \delta_{\mu\nu} - \partial_\mu \partial_\nu \right) {1\over
(x-y)^4} \left[ c_0 + \sum_{n=1} c_n \left( \ln M^2
(x-y)^2\right)^n\right]\ \ .\eqno(8.1)$$
If subdivergences are absent, $c_n = 0$ for $n\ge 1$, then one obtains the
conformal invariant form (1.6).  In other words, conformal invariance gives no
information beyond gauge invariance and cancellation of subdivergences.  We
expect that this last condition holds through two-loop order (for currents or
external gluons), and that logarithmic corrections beyond that are determined
by the renormalization group equations.

For vertex functions of vector fields or conserved currents, conformal
symmetry gives restrictions well-beyond gauge and scale invariance, requiring
that amplitudes are combinations of the tensors $C^{\rm sym}_{\mu\nu\rho}$
and $D^{\rm sym}_{\mu\nu\rho}$ of (1.10).  Any such combination obeys a simple
Ward identity of the form (1.4) or (2.10), and this suggests that the
background field formalism is relevant.  However, a simple Ward identity is not
sufficient for conformal invariance, and we have found explicit examples of
gauge and scale invariant, but not conformal, tensors in our study of the
background field vertex in a general gauge.

Our most striking result is that the one-loop background field vertex in
Feynman gauge is conformal invariant, so that the quark, ghost, and gluon loop
contributions are each linear combinations of $C^{\rm sym}_{\mu\nu\rho}$ and
$D^{\rm sym}_{\mu\nu\rho}$.  The practical consequence of the conformal
property is the relative ease of regularization, essentially because only
$D^{\rm sym}_{\mu\nu\rho}$ has an ultraviolet divergent Fourier transform.  It
is not difficult to show that the renormalized Ward identity can be satisfied
by adjustment of mass scale parameters,  $M_V$ in the regularized form of
$D^{\rm sym}_{\mu\nu\rho}$ and $M_\Sigma$ in the self-energy. Calculation of
the $\beta$-function is also a simple matter.

One may also study three-gluon vertex
functions with one or more external quantum gluons, of which some are required
for higher-loop computations in the background field formalism.  Such vertex
functions satisfy Slavnov--Taylor identities which are more complicated than
the simple Ward identity (1.4), and essentially for this reason one can
 rule out conformal
invariance.  Nevertheless, one can show easily that at one-loop order
the regulated form of these vertex functions can be expressed as a multiple of
the regulated $D^{\rm sym}_{\mu\nu\rho}$ plus a remainder which is ultraviolet
convergent.  A complete one-loop study of the differential regularization of
these vertices, which includes the vertex function of conventional
(non-background) gauge field theory, is an open problem whose solution should
be facilitated by the observation above.

Another open problem is the possible conformal property of the background
field four-gluon correlator, which is related to the three-gluon vertex by a
Ward identity.  Conformal symmetry is not a very restrictive property for a
four-point function.  Nevertheless, it would be useful to know whether the
basic primitive divergent one-loop amplitudes with two, three, or four
external background gluons share the property of conformal invariance for
non-coincident points in Feynman gauge.  One can predict that the quark and
ghost loop contributions to the four-point function are conformal invariant.

At present we do not have a real explanation of the gauge specific conformal
property we found, nor do we know that it has any significance beyond the
technical virtue of ease of regularization.  Further exploration of the
conformal scenario outlined in Section~VII may illuminate such questions.
\goodbreak
\bigskip
\centerline{\bf ACKNOWLEDGEMENTS}
\medskip
The authors thank J. I. Latorre who generously wrote the REDUCE program which
was useful at several points in this work.  (N.R.) is indebted to the MEC
(Spain) for a Fulbright Scholarship.
\vfill
\eject
\centerline{\bf APPENDIX A}
\medskip
\centerline{\bf DEVIATION FROM THE FEYNMAN GAUGE}
\medskip
It is natural to ask whether the conformal properties found for the background
self-energy and vertex are valid for general values of the gauge parameter
$a$.  A complete calculation, using the propagator (2.14) and the interaction
vertices ${\cal L}_1,\ldots,{\cal L}_4$ of (2.13) is feasible for the
self-energy and it gives exactly the same result as the Feynman gauge
calculation of (7.6) and is thus
independent of $a$.  This is of course to be expected,
since the one-loop
beta-function does not depend on the gauge parameter.

An analytic
calculation of the three-gluon vertex in a general gauge is very tedious
and we chose to study whether conformal invariance is preserved for a small
deviation from Feynman gauge. To this purpose one can expand the functional
$\Omega[B,J]_{j=0}$ in a series in $a$
$$\Omega[B,J] = \Omega[B,J]\bigg|_{a=1} + {\partial\Omega[B,J]\over \partial
a} \bigg|_{a=1} (a-1) + \ldots \eqno(\hbox{A.1})$$
and then analyze the conformal properties of the first coefficient of the
expansion ${\partial\Omega\over\partial a}\bigg|_{a=1}=\Omega^{(1)}$.  Such a
coefficient coincides with the vacuum expectation value of the gauge fixing
term and by means of a standard Ward-identity [Abers--Lee] of Yang--Mills
theories it is possible to relate it to other Green's functions which can be
more easily computed.

One can easily adapt to the background field formalism, the derivation of the
functional Slavnov--Taylor identity given by Abers and Lee.$^{18}$  The
result is
$$\eqalign{\int &db\det M\exp - \left\{ S[B+b] - S[B] + S_{gf}[b] + \int d^4x\,
J^a_\mu b^a_\mu\right\} \cr
&\cdot \left\{ {1\over a} G^a [B,b](x) + \int d^4y\, J^b_\mu(y) D^y_\mu [B+b]
\left[ M^{-1} (y,x)\right]^{ba}\right\} = 0 \cr}\eqno(\hbox{A.2})$$
where $G^a[B,b]$ is the gauge fixing functional, and
we must choose $J^a_\mu = D_\nu B^a_{\nu\mu}$ as in Section II, in order
to eliminate the linear ``tadpole'' terms in $S[B+b]-S[B]$.  Equation (A.2)
can be rewritten in terms of ghost fields as the functional ``expectation
value''
$$\ll {1\over a} D_\mu b^a_\mu(x) + f^{bcd} \int d^4y\, J^b_\mu (y) b^c_\mu(y)
 c^d(y) \bar{c}^a(x) \rr = 0\ \ ,\eqno(\hbox{A.3})$$
where we have used the covariant conservation of the current.  From (A.3)
it is now easy to obtain a Ward identity for $\Omega^{(1)}[B,J]$ by taking the
$D^x_\nu$ derivative, then a variation with respect to $J^a_\nu(x)$ and
finally integrating in $d^4x$, we arrive at
$$\eqalign{\Omega^{(1)}\left[B,D_\nu B_{\nu\mu}\right]
&=-{1\over 2} \int d^4x \ll
D_\mu b^a_\mu D_\nu b^a_\nu\rr \cr
&= -{1\over 2} f^{bcd} \int d^4x\, d^4y\, D_\rho B^b_{\rho\mu}(y) \ll b^c_\mu
(y) c^d(y) D^x_\nu \bar{c}^a(x) b^a_\nu(x)\rr \ \ .\cr}\eqno(\hbox{A.4})$$
in which it is understood that only connected graphs are included on the
right-hand side.  The linear deviation of the gluon vertex can now be
obtained from all graphs contributing to the third variational derivative of
(A.4) with respect to $B$.  (The second variational derivative vanishes
because the bilinear part of the effective action is the same in any gauge).

We focus our attention on the triangle diagrams with three external fields and
neglect seagull diagrams, because the triangles will give us a sufficient
condition to disprove conformal invariance for this linear deviation from the
Feynman gauge. These triangle diagrams are obtained from the Wick
contractions of the quantum fields in the double-integral term in (A.4) with
the vertices ${\cal L}_1$, ${\cal L}_3$ and ${\cal L}^g_i$ of the interaction
Lagrangian.  In several diagrams there is an effective two-point vertex from
the fields $\partial_\nu\bar{c}^a(x) b^a_\nu(x)$ in (A.4), and this leads
to the integral
$$\int d^4x {1\over (x-y)^2}\partial^x_\mu {1\over (x-z)^2} = 2\pi^2
{(z-y)_\mu\over (z-y)^2}\ \ .\eqno(\hbox{A.5})$$
One can show that the sum of all graphs gives the following contribution to
$\Omega^{(1)}_3$:
$$\eqalign{-&{1\over 2\left( 4\pi^2\right)^3} Cf^{abc} \int d^4x\, d^4y\, d^4z
\Biggl[ B^a_\mu(x) \partial^y_\rho B^{'b}_{\rho\nu}(y)
B^{'c}_{\nu\sigma}(z)\cr
&\cdot {(y-z)_\sigma\over (y-z)^2} \ {1\over (y-x)^2}
\upleftrightarrow{\partial}^x_\mu {1\over (x-z)^2} - B^{'a}_{\mu\nu} (x)
\partial^y_\rho B^{'b}_{\rho\sigma}(y) B^{'c}_{\sigma\mu}(z) {1\over (y-z)^2
(z-x)^2}\ {(y-x)_\nu\over (y-x)^2}\Biggr] \cr}\eqno(\hbox{A.6})$$
where $B^{'a}_{\mu\nu}(x) = \partial_\nu B^a_\nu - \partial_\nu B^a_\mu$ is
the linearized field strength.  This quantity is ultraviolet finite due to the
presence of three external derivatives.
One might then suspect that if (A.6) is conformal invariant, its variational
derivative with respect to $B^a_\mu(x)$, $B^b_\nu(y)$, $B^c_\rho(z)$ would be
proportional to the ultraviolet convergent tensor in $C^{\rm
sym}_{\mu\nu\rho}$.  However, partial integration of the three external
derivatives produces both genuine ``triangular terms'' depending on $(x-y)^2$,
$(y-z)^2$ and $(z-y)^2$ and ``semi-local terms'' containing $\delta(z-y)$ {\it
etc.\/} (which we drop here).  The triangular part could then be a combination
of $C^{\rm sym}_{\mu\nu\rho}$ and $D^{\rm sym}_{\mu\nu\rho}$ with the
divergent part of $D^{\rm sym}_{\mu\nu\rho}$ cancelled by the neglected
semi-local terms and seagull graphs.

After partial integration of the external derivatives and tedious algebra to
simplify the result, we find that (A.6) can be rewritten as
$$\eqalign{&-{4C\over \left( 4\pi^2\right)^3} f^{abc} \int d^4x\, d^4y\, d^4z\,
B^a_\mu (x) B^b_\nu(y) B^c_\rho(z) \cr
&\times \Biggl\{ \delta_{\mu\nu} (x-z)_\rho \Biggl[ {(x-z)^2 +
(y-z)^2+(x-y)^2\over (x-y)^4 (z-y)^4 (x-z)^4} \cr
&- {6\over (z-x)^4(x-y)^6} +
{2\over (z-y)^4 (x-y)^6}\cr
&- {4\over (x-y)^4 (z-x)^6} + {4(z-y)^2\over (z-x)^6(x-y)^6}\Biggr] \cr
&- 2{(x-z)_\mu\over (x-z)^4}\ {(x-y)_\nu\over (x-y)^4}\ {(y-z)_\rho\over
(y-z)^4} - 8{(x-y)_\mu (z-y)_\nu (x-z)_\rho\over (x-y)^6 (y-z)^6} \cr
&- 4{(x-y)_\mu (x-z)_\nu (x-z)_\rho\over (x-y)^6 (x-z)^2 (z-y)^2} \left[
{1\over
(z-y)^2} + {1\over (x-z)^2}\right] \cr
&+ 4{(x-z)_\mu (z-y)_\nu (x-z)_\rho\over (z-y)^6 (x-y)^2 (x-z)^2} \left[
{1\over (x-y)^2}+ {1\over (x-z)^2} \right]\Biggr\}\ \ .\cr}\eqno(\hbox{A.7})$$
As a check of the computations one can show that (A.7) is gauge invariant.
Nevertheless, close inspection shows that its (symmetrized) variational
derivative
cannot be expressed in terms of
the conformal tensors $C^{\rm sym}_{\mu\nu\rho}$ and $D^{\rm
sym}_{\mu\nu\rho}$ and consequently is
not conformal invariant.  Both analytic calculation and symbolic manipulation
confirm this, so we can conclude that for small deviations from
the Feynman gauge the conformal properties of the three-gluon vertex are lost.
\vfill
\eject
\centerline{\bf APPENDIX B}
\medskip
\centerline{\bf MIXED THREE GLUON VERTICES}
\medskip
In this Appendix we discuss further the structure of one-loop vertex functions
with both background and quantum external gluons.  An argument was given in
Section~III that these vertices are not conformal invariant, but they can be
expressed as a multiple of the conformal tensor $D^{\rm sym}_{\mu\nu\rho}$
plus an ultraviolet finite remainder.  Our explicit real space computations
support this picture, but we do not give full details here, since our main
concern is to show that the coefficient of the $D^{\rm sym}_{\mu\nu\rho}$
leads, after regularization of this tensor, to a renormalization scale
dependence in agreement with the work of Kluberg--Stern and Zuber.$^{11}$
In our notation, the result of these authors for the renormalized action
involving three-gluon vertices in the background field formalism is:
$$S^R = \int d^4x \left\{ {1\over g^2 Z^2_g} {\cal L}_{\rm YM} \left( B + gZ_g
Z^{1/2}_3 b\right) + {1\over 2a} \left( D_\mu b^a_\mu\right)^2 \right\}
\eqno(\hbox{B.1})$$
where $g_0 = gZ_g$ is the bare coupling constant and $Z_3$ is the wavefunction
renormalization factor of the quantum gauge field, $b^a_\mu(x)_{\rm
bare}=Z^{1/2}_3 b^a_\mu(x)$.  In the second term, one sees that
renormalization of the gauge fixing parameter $a$ is also needed, with $a_0 =
Z_aa$ and $Z_a=Z_3$.

{}From Eq.~(B.1) we see that the counterterm for the overall divergence of any
three-gluon vertex has the form of the bare Yang--Mills vertex.  In
differential regularization this implies that the part of the one-loop
amplitudes which requires regularization is a numerical coefficient times the
singular part of the tensor $D^{\rm sym}_{\mu\nu\rho}(x,y,z)$.  Indeed, this
is the only possibility for vertices with three background or three quantum
gluons, because of Bose symmetry,  but the result is not obvious for mixed
vertices.

The Feynman rules for the quantum gluon vertices are given by ${\cal L}_q$ and
${\cal L}'{}^g$ in Eqs.~(2.12) and (2.16).  It is straightforward to apply
differential regularization to the 1PI diagrams which contribute to the
one-loop amplitude for the mixed vertices, and identify the coefficient of the
scale dependent term which is proportional to the semi-local scale dependent
part of the regulated tensor $D^{\rm sym}_{\mu\nu\rho}$ in (4.7).

In the mixed vertex with two background and one external quantum gluon, both
ghost seagull graphs vanish separately, due to the antisymmetry of the group
structure constants $f^{abc}$.  The contribution of the two gluon seagull
diagrams is just the divergent part of $D^{\rm sym}_{\mu\nu\rho}(x,y,z)$ times
a numerical coefficient, while the triangle graphs yield a multiple of $D^{\rm
sym}_{\mu\nu\rho}(x,y,z)$ plus an ultraviolet finite piece.  We give here only
the mass scale dependence of the final result
$$M{\partial\over\partial M} \ {\delta^3\Omega[B,J]_{j=0} \over
\delta j^a_\mu(x) \delta B^b_\nu(y) \delta B^c_\rho(z)} = g\,f^{abc} {C\over
\left(4\pi^2\right)^3}\  {32\over 3} M {\partial\over\partial M} D^{\rm
sym}_{\mu\nu\rho}(x,y,z) \ \ .\eqno(\hbox{B.2})$$

In the mixed vertex with one background and two quantum gluons the
contributions from all diagrams combine to yield a multiple of the divergent
part of $D^{\rm sym}_{\mu\nu\rho}(x,y,z)$, plus an ultraviolet finite
remainder from the triangle graphs.  The formal scale dependence of the
one-loop vertex is
$$M{\partial\over \partial M}\ {\delta^3\Omega [B,J]_{j=0}\over\delta
j^a_\mu(x) \delta j^b_\nu (y) \delta B^c_\rho(z)} = g^2\, f^{abc} {C\over
\left( 4\pi^2\right)^3} \ {20\over 3} M {\partial\over \partial M} D^{\rm
sym}_{\mu\nu\rho} (x,y,z)\ \ .\eqno(\hbox{B.3})$$
The renormalized action given by Eq.~(B.1) suggests that the renormalization
group equation for a three-gluon vertex function with $n_b$ external quantum
gluons, $\Gamma^{(n_b)}_{\mu\nu\rho}$, takes the form
$$\left[ M {\partial\over \partial M} + \beta(g) {\partial\over \partial g} +
\delta(g) a {\partial\over \partial a} - n_b \gamma(g) \right]
\Gamma^{(n_b)}_{\mu\nu\rho} = 0 \eqno(\hbox{B.4})$$
where $\gamma(g) = g^2 \gamma_1 + g^4 \gamma_2 + \ldots$ in the notation of
Section~VII.  We have already discussed the three background gluon vertex in
Section~VI, where we computed the one-loop coefficient of $\beta(g)$ by means
of the renormalization group equation (7.4).

Then we can apply Eq.~(B.4) to the vertex with two background and one external
quantum gluons.  We combine the classical term of order $1/g$ with the
one-loop result (B.2) and use the scale derivative of the regulated tensor
$D^{\rm sym}_{\mu\nu\rho}$, given by Eq.~(6.3).  Since the gauge fixing action
does not contain $BBb$ terms, the $\delta(g) a{\partial\over\partial a}$ term
does not contribute to lowest order.  In the final result to order $g$,
$M{\partial\over\partial M}$ acting on the one-loop amplitude is balanced by
$g^3\beta_1 {\partial\over\partial g} - g^2\gamma_1$ applied to the classical
term.  Using the value of $\beta_1$ in (6.4) we obtain the anomalous dimension
in the Feynman gauge
$$\gamma_1 = - {5\over 48\pi^2} C\eqno(\hbox{B.5})$$
which agrees with the known result.$^3$

Finally we apply a similar test of the renormalization group Eq.~(B.4) to the
combined classical and one-loop $Bbb$ vertex.  Here the $\beta(g){\partial
\over\partial g}$ term does not contribute to order $g^2$, and the scale
derivative is balanced by $g^2 \delta_1 a{\partial\over\partial a} -
2g^2\gamma_1$ applied to the classical term.  The result confirms that
$\delta_1 = - 2\gamma_1$ which agrees with the general argument that
$\delta(g) = - 2\gamma(g)$ (which follows from $Z_a = Z_3$), and also with the
relation found from the study of $bb$ two-point function in Section~VII.

Using these results for $\beta_1$ and $\gamma_1$, the renormalization group
equation for the three quantum gluon vertex function to one-loop implies that
its mass scale dependence in the Feynman gauge must be
$$M {\partial\over \partial M} \  {\delta^3\Omega [B,J]_{j=0}\over \delta
j^a_\mu (x) \delta j^b_\nu (y) \delta j^c_\rho(z)} = g^3\,f^{abc} {C\over
\left( 4\pi^2\right)^3} \  {8\over 3} M {\partial\over\partial M} D^{\rm
sym}_{\mu\nu\rho}(x,y,z)\ \ .\eqno(\hbox{B.7})$$
In this case we have not performed the explicit graphical computation to test
this result.
\vfill
\eject
\centerline{\bf APPENDIX C}
\medskip
\centerline{\bf THREE-GLUON VERTEX IN A GENERAL GAUGE}
\medskip
It is clearly of interest to see if conformal invariance holds for any value
of the gauge parameter $a$ different from the Feynman gauge value $a=1$.  The
case of the Landau gauge ($a=0$) is especially interesting, because the
simplified renormalization group equations in this gauge permit exact conformal
invariance at the two-loop level as discussed in Section~VII.

The analytic calculation of the three-gluon vertex in a general gauge is
very tedious so we have used symbolic manipulation based on a core REDUCE
program written by J.~I.~Latorre.  It  uses symbolic logic to calculate
partial derivatives of any translation invariant tensor function
of three points $x_\mu$, $y_\nu$, $z_\rho$.

We compute analytically the Wick contractions for all triangle graphs
involving the trilinear interaction terms ${\cal L}_1$, ${\cal L}_3$,
${\cal L}_4$ of Eq.~(2.13), using the general gauge propagator (2.14).  The
contribution to the cubic part of the effective action can be expressed as
$$\Omega^{ijk} [B] = {C\over \left( 4\pi^2\right)^3} f^{abc} \int d^4x\,
d^4y\, d^4z\, B^a_\mu (x) B^b_\nu(y) B^c_\rho(z) A^{ijk}_{\mu\nu\rho} (x,y,z)
\eqno(\hbox{C.1})$$
where $i,j,k$ indicates which vertices contribute to a given triangle.  The
program computes the derivatives in $A^{ijk}_{\mu\nu\rho}(x,y,z)$ and then
adds the permutations that are necessary to produce the fully permutation odd
contributions $S^{ijk}_{\mu\nu\rho}(z,y,z)$ to the three-gluon vertex
function.  Each $S^{ijk}_{\mu\nu\rho}(x,y,z)$ is a cubic function of the
gauge parameter $a$.  It is easily seen that possible $1/a$ poles from the
interaction vertex ${\cal L}_4$ cancel, because ${\cal L}_4$ always generates
the divergence of the propagator
$$\partial^x_\mu \ll b_\mu(x) b_\nu(y) \rr = a {1\over 4\pi^2} \partial^x_\mu
{1\over (x-y)^2} \eqno(\hbox{C.2})$$
which cancels the $1/a$ singularity.  As explained in Appendix A, we neglect
both seagull graphs and $\delta$-functions in the triangle graphs because they
are not needed to test the conformal property for $x\not=y\not=z$.

We express the full vertex amplitude as a series in $(a-1)$:
$$J^0_{\mu\nu\rho} + (a-1) J^1_{\mu\nu\rho} + (a-1)^2 J^2_{\mu\nu\rho} +
(a-1)^3 J^3_{\mu\nu\rho} \eqno(\hbox{C.3})$$
and we perform the following consistency checks of the computation.
\medskip
\item{1)}Each term $J^i_{\mu\nu\rho}(x,y,z)$ satisfies the divergenceless
property
$${\partial\over \partial z_\rho} J^i_{\mu\nu\rho} (x,y,z) = 0 \ \ ,\qquad
i=0,\ldots 3 \ \ ,\eqno(\hbox{C.4})$$
which is required since the Ward identity (2.10) vanishes for non-coincident
points $x\not=y\not=z$.
\medskip
\item{2)}The $J^0_{\mu\nu\rho}$ piece agrees with the Feynman gauge result we
obtained analytically in Section~III.
\medskip
\item{3)}The $J^1_{\mu\nu\rho}$ term agrees with the analytic calculations of
linear deviation from Feynman gauge (see Appendix~A).
\medskip
\noindent These highly non-trivial checks give confidence in the computer
result, so we go on to analyze the conformal properties of the tensors
$J^i_{\mu\nu\rho}(x,y,z)$ in Eq.~(C.3).  $J^0_{\mu\nu\rho}(x,y,z)$ is
conformal invariant, as we already knew since it is the only contribution when
$a=1$.  For the coefficient of the $(a-1)^3$ term in (C.3) we find the
conformal tensor
$$J^3_{\mu\nu\rho}(z,y,z) = {1\over 8} C^{\rm sym}_{\mu\nu\rho}(x,y,z) \ \
.\eqno(\hbox{C.5})$$

This result may be easily explained, since the piece cubic in the gauge
parameter $a$ involves only the term $a\partial_\mu\partial_\nu\ln
(x-y)^2$ in the quantum gluon propagator, which has the correct conformal
inversion property, and propagators are connected for large $a$ only by
vertices from the Yang--Mills action $S[B+gb]$, which is conformal invariant.
This implies that $J^3_{\mu\nu\rho}$ must be a linear combination of $C^{\rm
sym}_{\mu\nu\rho}$ and $D^{\rm sym}_{\mu\nu\rho}$.

Finally, we analyze the linear and quadratic terms in Eq.~(C.3).
$J^1_{\mu\nu\rho}(x,y,z)$ has the non-conformal structure given in Eq.~(A.7)
and $J^2_{\mu\nu\rho}(x,y,z)$ is a far more complicated expression, so we
chose to study the structure of these tensors in the limit as one of the
points $(y_\nu)$ goes to infinity.  In this limit, both conformal tensors
$C^{\rm sym}_{\mu\nu\rho}$ and $D^{\rm sym}_{\mu\nu\rho}$ have the form
$${c_1\over y^6} \left( \delta_{\nu\sigma} - {2y_\nu y_\sigma\over y^2}
\right) {1\over x^4} \left\{ \delta_{\mu\rho} x_\sigma - \delta_{\mu\sigma}
x_\rho - \delta_{\rho\sigma} x_\mu + c_2 {x_\mu x_\rho x_\sigma\over x^2}
\right\} \eqno(\hbox{C.6})$$
where we have set $z=0$ for simplicity and the coefficients $c_1$ and $c_2$
depend on the conformal tensor ($c_1=8$, $c_2=4$ for $C^{\rm
sym}_{\mu\nu\rho}$; $c_1 = -4$, $c_2 = -2$ for $D^{\rm sym}_{\mu\nu\rho}$).

We study only the terms containing $\delta_{\mu\rho}$ in $J^1_{\mu\nu\rho}$
and $J^2_{\mu\nu\rho}$, because they give us a sufficient condition to
disprove conformal invariance away from the Feynman gauge.  The leading term
as $y_\nu$ goes to infinity is of order $1/y^4$, which is absent in the
conformal tensor structure, but it has the same form in both tensors
$J^1_{\mu\nu\rho}$ and $J^2_{\mu\nu\rho}$ and therefore it can be eliminated
in a suitable combination of them, namely $J^1_{\mu\nu\rho} -8
J^2_{\mu\nu\rho}$.  However, when we study the next-to-leading order, which is
$1/y^6$, we see that this linear combination of $J^1_{\mu\nu\rho}$ and
$J^2_{\mu\nu\rho}$ does not have the correct structure of a conformal tensor
given by (C.6).  We can then conclude that no linear combination of
$J^1_{\mu\nu\rho}$ and $J^2_{\mu\nu\rho}$ can be a combination of conformal
tensors and therefore among $D_\mu b^a_\mu$ background gauges only in the
Feynman gauge the one-loop three-gluon vertex function is conformal invariant.
\vfill
\eject
\centerline{\bf REFERENCES}
\medskip
\item{1.}D. Z. Freedman, K. Johnson and J. I. Latorre, {\it Nucl. Phys.\/}
{\bf B371} (1992) 329.
\medskip
\item{2.}P. E. Haagensen, {\it Mod. Phys. Lett. A\/}, in press.
\medskip
\item{3.}W. Marciano and H. Pagels, {\it Phys. Rep.\/} {\bf 36} (1978) 137.
\medskip
\item{4.}B. S. DeWitt, {\it Phys. Rev.\/} {\bf 162} (1967) 1195, 1239.
\medskip
\item{5.}G. 't~Hooft, {\it Nucl. Phys.\/} {\bf B62} (1973) 444.
\medskip
\item{6.}L. F. Abbott, {\it Nucl. Phys.\/} {\bf B185} (1981) 189.
\medskip
\item{7.}L. F. Abbott, M. T. Grisaru and R. K. Schaefer, {\it Nucl. Phys.\/}
{\bf B229} (1983) 372.
\medskip
\item{8.}A. Rebhan, {\it Nucl. Phys.\/} {\bf B298} (1988) 726.
\medskip
\item{9.}E. J. Schreier, {\it Phys. Rev.\/} {\bf D3} (1971) 980.
\medskip
\item{10.}M. Baker and K. Johnson, {\it Physica\/} {\bf 96A} (1979) 120.
\medskip
\item{11.}H. Kluberg--Stern and J. B. Zuber, {\it Phys. Rev.\/} {\bf D12}
(1975) 482.
\medskip
\item{12.}B. S. DeWitt, in {\it General Relativity: An Einstein Centenary
Survey\/}, S. Hawking and W. Israel, eds. (Cambridge University Press,
Cambridge, UK, 1979); G. A. Vilkovisky, {\it Nucl. Phys.\/} {\bf B234} (1984)
125; A. Rebhan, {\it Nucl. Phys.\/} {\bf B288} (1987) 832.
\medskip
\item{13.}R. Jackiw, in {\it Field Theoretic Investigations in Current
Algebra\/}, S. B. Trieman, R. Jackiw, B. Zumino and E. Witten, eds. (World
Scientific, Singapore, 1985).
\medskip
\item{14.}R. Jackiw, {\it Phys. Rev. Lett.\/} {\bf 41} (1978) 1635; B.
de~Wit and D. Z. Freedman, {\it Phys. Rev.\/} {\bf D12} (1975) 2286.
\medskip
\item{15.}E. de~Rafael and J. L. Rosner, {\it Ann. Phys.\/} {\bf 82} (1974)
369.
\medskip
\item{16.}C. Itzykson and J. B. Zuber, {\it Quantum Field Theory\/}
(McGraw-Hill, New York, 1980).
\medskip
\item{17.}R. Mu\~noz--Tapia, X. Vilas\'{\i}s--Cardona,
D. Z. Freedman and K. Johnson,  manuscript in preparation.
\medskip
\item{18.}E. S. Abers and B. W. Lee, {\it Phys. Rep.\/}{\bf 9C} (1973) 1.
\vfill
\eject
\centerline{\bf FIGURE CAPTIONS}
\medskip
\item{Fig.~l:}1PI diagrams involving ghosts which contribute to the one-loop
three-gluon vertex.
\bigskip
\item{Fig.~2:}1PI diagrams involving fermion loops which contribute to the
three-gluon vertex.
\bigskip
\item{Fig.~3:}1PI diagrams involving gluon loops which contribute to the
three-gluon vertex.

\par
\vfill
\end